\documentclass[aps,prd,floatfix,amsmath,amssymb,preprint,eqsecnum,nofootinbib,superscriptaddress]{revtex4}
\usepackage{graphicx}
\usepackage{dcolumn}
\usepackage{bm}
\usepackage{epstopdf}
\usepackage{amsmath, amsfonts, amssymb}
\usepackage{pstricks}
\usepackage{amsxtra}
\usepackage{amsthm}

\newcommand{\beq}{\begin{equation}}
\newcommand{\eeq}{\end{equation}}
\newcommand{\ba}{\begin{array}{ccc}}
\newcommand{\ea}{\end{array}}
\newcommand{\nn}{\nonumber \\}
\def\bea{\begin{eqnarray}}
\def\eea{\end{eqnarray}}

\usepackage{setspace} 
\begin{document}
\preprint{NSF-KITP-11-259}
\preprint{arXiv:1112.0573}
\title{Hidden Fermi surfaces in compressible states\\ of gauge-gravity duality}
\author{Liza Huijse}
\author{Subir Sachdev}
\affiliation{Department of Physics, Harvard University, Cambridge MA
02138}
\affiliation{Kavli Institute for Theoretical Physics, University of California, Santa Barbara CA 93106-4030}
\author{Brian Swingle}
\affiliation{Department of Physics, Harvard University, Cambridge MA
02138}

\date{December 2, 2011 \\
\vspace{1.6in}}
\begin{abstract}
General scaling arguments, and the behavior of the thermal entropy density, are shown to lead to an infrared metric
holographically representing a compressible state with hidden
Fermi surfaces. This metric is characterized by a general dynamic critical exponent, $z$, and a specific
hyperscaling violation exponent, $\theta$. The same metric exhibits a logarithmic violation of the area law of entanglement entropy,
as shown recently by Ogawa {\em et al.} (arXiv:1111.1023).
We study the dependence of the entanglement entropy on the shape of the entangling region(s),
on the total charge density, on temperature, and on the presence of additional visible Fermi surfaces of gauge-neutral fermions;
for the latter computations, we realize the needed metric in an Einstein-Maxwell-dilaton theory.
All our results  support
the proposal that the holographic
theory describes a metallic state with hidden Fermi surfaces of fermions carrying gauge charges of deconfined gauge fields.
\end{abstract}

\maketitle

\section{Introduction}
\label{sec:intro}

Much recent work \cite{nernst,sslee0,denef0,hong0,zaanen1,hong1,denef,faulkner,pufuigor,polchinski,gubserrocha,hong2,kiritsis,kiritsis2,sean1,sean2,sean3,seanr,larus1,larus2,eric,kachru2,kachru3,kachru4,trivedi,zaanen2,ssffl,mcphys,liza,leiden,hong4,hong5,pp,waldram,yarom,ssfl,ssarcmp,larus3,lizasean,tadashi}
has focused on the holographic description of compressible metallic states of quantum matter.

From a condensed matter perspective, a key issue in the zero temperature theory of any metallic state is the fate
of the Luttinger relation \cite{liza}, equating the total charge density, $ \mathcal{Q} $ to the volumes enclosed
by the Fermi surfaces. An important feature of the early theories \cite{sslee0,hong0,zaanen1,hong1} of holographic metals was that the Luttinger relation
was badly violated: the holographic theory exhibits Fermi surfaces of gauge-neutral `mesinos' (in the terminology
of Ref.~\cite{lizasean}), but they enclose volumes which are much smaller than the charge density,
$\mathcal{Q}$, of the boundary theory. (In the AdS/CFT correspondence, $\mathcal{Q}$ diverges
with large numbers of colors, $N$,
while the mesino Fermi surfaces enclose volumes of order unity.) It was argued in Refs.~\cite{ssffl,liza} that this deficit
must be made up by {\em hidden\/} Fermi surfaces of `fractionalized' fermions carrying gauge charges; we will refer to these
gauge-charged fermions generically as `quarks' (although they need not co-incide with the elementary fermions of any weak-coupling
formulation of the boundary theory; in the context of supersymmetric gauge theories, we could also refer to them as `gauginos', but
we will not do so here).
The holographic theory is only able to access gauge-invariant observables, and so direct signatures of the quark Fermi surfaces remain hidden to probes on the boundary.

In subsequent holographic studies, states obeying the standard Luttinger relation were indeed
found \cite{sean2,sean3,hong4,ssfl}. However, they all had Fermi surfaces of gauge-neutral mesinos alone,
and are all ultimately expected to be confining Fermi liquids (FLs) at low temperatures.
Our interest here is primarily on non-zero density states in which the gauge field is deconfined.
Such states are expected to have Fermi surfaces of quarks alone, as in the `non-Fermi liquid'
(NFL) states of Ref.~\cite{liza}. Another deconfined state is the `fractionalized Fermi liquid' (FL*) \cite{ffl1,ffl2},
which has co-existing quark and mesino Fermi surfaces \cite{liza}; holographic phases with charge fractionalization were considered in \cite{lizasean}. The key problem of holographically detecting the hidden quark Fermi surfaces in these deconfined states
remained open.

In a recent paper, Ogawa, Takayanagi, and Ugajin \cite{tadashi} have proposed an elegant solution to this conundrum: compute the
holographic entanglement entropy of the compressible state, and match it to that expected from the hidden quark Fermi surfaces.
They considered a class of Einstein-Maxwell-dilaton (EMD) theories of compressible quantum states \cite{gubserrocha,kachru2,kachru3,trivedi,kiritsis,kiritsis2,larus1},
and described conditions under which the area law of entanglement was logarithmically violated.
The quark Fermi surfaces are expected \cite{swingle,tarun,seidel} to display such a logarithmic violation of the area law,
and so such EMD theories can be regarded as effective holographic descriptions of deconfined compressible states with quark Fermi surfaces,
even though the EMD theories have no explicit fermionic degrees of freedom. Indeed, even in quantum models with only 
bosonic degrees of  freedom, non-superfluid compressible phases at non-zero $\mathcal{Q}$ are expected to have Fermi surfaces of fractionalized fermions \cite{motfish}.

\subsection{Violation of hyperscaling, entropy, and entanglement entropy}

Here we present a separate and simple general scaling argument for the infrared (IR) metric
of holographic theories of non-Fermi liquid states with hidden Fermi surfaces
in $d$ spatial dimensions. Condensed matter physicists can find an introduction to the 
relevant concepts of the holographic method in Ref.~\cite{ssarcmp}.
We can write an arbitrary, holographic infrared (IR) metric in the scaling form
\beq
ds^2 = \frac{1}{r^2} \left( - \frac{dt^2}{r^{2d (z-1)/(d-\theta)}} + r^{2\theta/(d-\theta)} dr^2 + d x_i^2 \right) \label{zmetric}
\eeq
where $t$ is time, $x_i$ ($i = 1 \ldots d$) are the $d$ spatial directions, and $r$ is the emergent holographic
direction. All numerical pre-factors (which are independent of $t$, $x_i$, $r$) have been set equal to unity,
and we have used reparametrization invariance in $r$ to fix the co-efficient of $dx_i^2$ to equal $1/r^2$.
Such metrics can be realized for compressible phases at
generic charge densities $\mathcal{Q}$, and the numerical prefactors then depend upon $\mathcal{Q}$ in a manner which will be discussed
in Section~\ref{sec:EMD}.
The boundary quantum theory is at $r=0$ (where the form (\ref{zmetric}) will not apply),
and its low energy physics is captured by the IR limit, $r \rightarrow \infty$.
This metric is characterized by two independent exponents, $z$ and $\theta$, which have been chosen so that scale transformations
by a factor $\zeta$ have the simple form
\bea
x_i &\rightarrow& \zeta \, x_i \nn
t &\rightarrow& \zeta^z \, t \nn
ds &\rightarrow & \zeta^{\theta/d} \, ds . \label{scaling}
\eea

This transformation makes it clear that $z$ is the {\em dynamic critical exponent\/}.
Relativistic conformal field theories have $z=1$ and $\theta = 0$.
Theories with $\theta=0$ but $z\neq 1$ are often referred to as ``Lifshitz'' theories \cite{mulligan,horowitz}.

The novel feature of the metric (\ref{zmetric}) is that the proper distance $ds$ of the emergent spacetime
transforms non-trivially under scale transformations, with the exponent $\theta$.
In the usual AdS/CFT correspondence, the proper distance is invariant with $\theta =0$, rather
than transforming covariantly under scale transformations. There is a natural connection between volume elements
in the holographic space, and various entropic measures of the boundary theory, and so this suggests that
a non-zero $\theta$ will modify the scale transformation of the thermal entropy density, $S$. Indeed, using this reasoning we
now show that $\theta$ is the {\em hyperscaling violation exponent\/} \cite{dsf} for the boundary theory;
thus, non-invariance of the proper distance in the holographic theory implies violations of hyperscaling on the boundary.
Hyperscaling is the property that the free energy scales by its naive dimension;
specifically, in theories with hyperscaling $S$, has the temperature, $T$, dependence $S \sim T^{d/z}$ \cite{fwgf,book}. In
the present holographic context, we can determine $S$ by its proportionality
to the area of the horizon of a black brane which appears for $T>0$;
so for a horizon at $r=r_h$, we have from (\ref{zmetric}) that $S \sim r_h^{-d}$. Also, from
and (\ref{zmetric}) and (\ref{scaling}) we determine that $r$ scales as
\beq
r \rightarrow \zeta^{(d-\theta)/d} r, \label{rscaling}
\eeq
and so we deduce that $r^d$ scales as
$t^{(d-\theta)/z}$. Noting that $T$ is an inverse time, we can conclude that
\beq
S \sim T^{(d-\theta)/z}, \label{st1}
\eeq
establishing that $\theta$ is the hyperscaling violation exponent;
an explicit computation of the entropy density obeying
(\ref{st1}) will appear in Section~\ref{sec:temp}, and analogous computations have appeared in
other works \cite{gubserrocha,trivedi,kiritsis,kiritsis2}.
We are interested here in compressible states with fermionic excitations which are gapless
on a $(d-1)$-dimensional surface in momentum space (the Fermi surface),
and which disperse along the single dimension transverse to the surface with dynamic critical exponent $z$;
such states should have an entropy density
\beq
S \sim T^{1/z}, \quad \mbox{for general $d$;} \label{st2}
\eeq
the entropy is that of the transverse excitations at each Fermi surface point, times the area of the Fermi surface.
This is also the behavior found in recent
gauge theories \cite{motfish,leen,metnem,mross} of non-Fermi liquid states in $d=2$, provided we identify $z$ with the dynamic scaling of the
Green's function of the gauge-charged fermions in the direction normal to the Fermi surface (in Ref.~\cite{metnem}, the symbol $z$ was used for the dynamic scaling
of a boson Green's function).
So such compressible states must have
\beq
\theta = d-1. \label{theta}
\eeq
Eq.~(\ref{theta}) is a key constraint on holographic theories of compressible states with hidden Fermi surfaces, and our main results will be restricted to systems with IR metrics which obey (\ref{zmetric}) and (\ref{theta}).

The arguments above also show that we can interpret $\theta$  as the dimension of the
momentum-space surface on which
there are zero-energy excitations. The value (\ref{theta}) is generic for systems with Fermi surfaces.

\begin{figure}[tbp]
  \centering
  \includegraphics[width=4in]{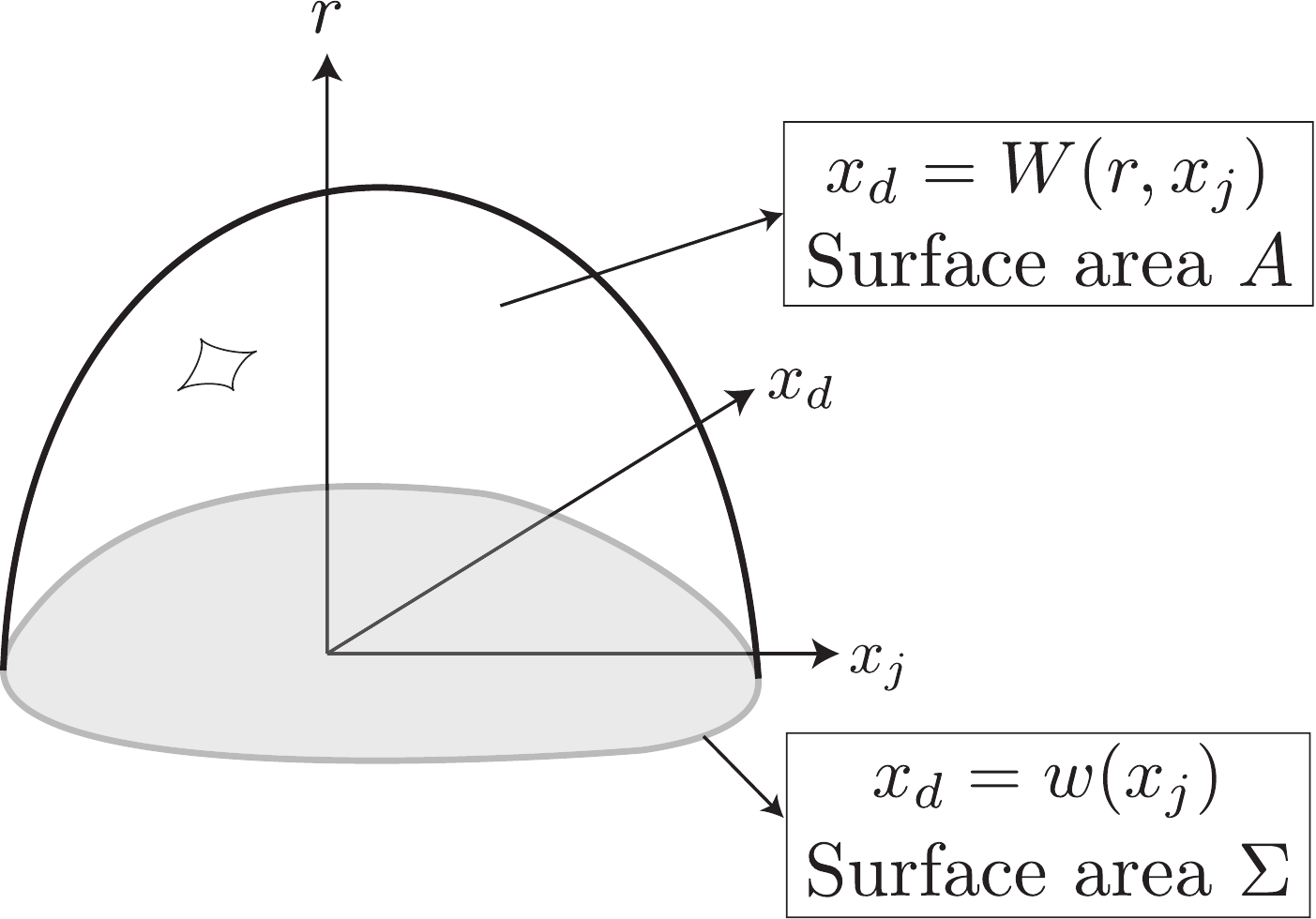}
  \caption{Geometry of holographic entanglement. The $d$ spatial co-ordinates are
  $x_i \equiv (x_j, x_d)$, with $j=1,\ldots,d-1$, and the emergent holographic direction is $r$.
  The entangling region of the compressible quantum state is shown shaded;
  its boundary is the surface described in co-ordinate patches by $x_d = w(x_j)$, $j=1,\ldots,d-1$, which has surface area $\Sigma$.
  This boundary is extended along the holographic
  direction into the surface written locally as $x_d = W (r, x_j)$; the holographic entanglement \cite{rt} is proportional to
  the surface area, $A$, of this extended boundary. The area $A$ has to be minimized while keeping the surface
  $x_d = w (x_j)$ fixed.
  }
  \label{fig:surface}
\end{figure}

We will compute the holographic entanglement entropy \cite{rt},
$S_{E}$, for theories obeying (\ref{zmetric}) and (\ref{theta}) for {\em arbitrary\/}
smooth shapes of the entangling region(s), and general $T$ in Section~\ref{sec:shape}.
At $T=0$, and for a single connected entangling region as in Fig.~\ref{fig:surface},
we find a logarithmic violation of the area law for $\theta=d-1$, and to leading log accuracy
\beq
S_{E} = \eta \, \mathcal{Q}^{(d-1)/d} \Sigma \, \ln \biggl(\mathcal{Q}^{(d-1)/d} \Sigma \biggr), \label{see}
\eeq
in the regime where the argument of the logarithm is large, and so $S_{E}$ can be computed entirely from scales
where (\ref{zmetric}) applies. Here
 $\eta$ is a dimensionless numerical
constant which depends upon the couplings of the holographic theory, but is {\em independent\/} of $\mathcal{Q}$ and
of any property of the entangling region.
The only dependence on the entangling region is via $\Sigma$, which is its $(d-1)$-dimensional surface area
(in $d=2$, $\Sigma$ is the perimeter of the entangling region---see Fig.~\ref{fig:surface}).
In general, gapless theories have an entanglement entropy which
depends upon the full geometry of the entangling region \cite{rt,cardy,mfs},
and not just on the surface area $\Sigma$. However, systems with
gapless fermionic excitations on a $(d-1)$-dimensional spherical Fermi surface \cite{klich,wolf,barthel,haas1,vicari,swingle,tarun,seidel}
have an entanglement entropy which has
precisely the form of Eq.~(\ref{see}), depending {\em only\/} on $\Sigma$ and no other characteristic of the smooth
entangling region. 

The specific $\mathcal{Q}$-dependence of (\ref{see}) arises from the $\mathcal{Q}$-dependence of the numerical prefactors
of (\ref{zmetric}). The latter dependence is quite complicated, and depends on details of the ultraviolet (UV) physics.
However, all of this UV dependence cancels out in the entanglement entropy, and only the universal 
IR $\mathcal{Q}$-dependence shown
in (\ref{see}) remains; this 
key result is established in Section~\ref{sec:EMD}.
Upon using the Luttinger relation, $\mathcal{Q} \sim k_F^{d}$, between
the charge density and the Fermi wavevector $k_F$, the $k_F$ dependence in (\ref{see}) is also identical to that
found in the Fermi surface computations \cite{klich,wolf,barthel,haas1,vicari,swingle,tarun,seidel}.
Thus, remarkably, the Luttinger relation of condensed matter physics is connected to 
some of the central principles of the holographic theory: Gauss's law \cite{ssfl}, the area law
for entanglement entropy \cite{rt}, and the universal $\mathcal{Q}$ dependence of the metric
of the holographic space.

We also examine the nature of entanglement between disjoint regions (the ``mutual information''),
and again find agreement with the Fermi surface results.

All of the above features of (\ref{st2}) and (\ref{see}) support the hidden Fermi surface interpretation of holographic
theories with the metric (\ref{zmetric}) for $\theta = d-1$.

We will also extend the EMD theories by adding explicit
fermionic degrees of freedom in Section~\ref{sec:fermions}.
These fermions are gauge-neutral mesinos, and form Fermi surfaces which are
directly visible in the holographic theory. In the language of Ref.~\cite{liza}, we are moving from the NFL state
to a FL* state with co-existing quark (hidden) and mesino (visible) Fermi surfaces.
We find that the FL* state also has the holographic entanglement entropy (\ref{see}), but with
$\mathcal{Q} \rightarrow \mathcal{Q} - \mathcal{Q}_{\rm mesino}$.
The FL* state has $\mathcal{Q}_{\rm quark} = \mathcal{Q} - \mathcal{Q}_{\rm mesino}$, and so this result
is consistent with the holographic entanglement entropy being a measure of the hidden quark Fermi surfaces.

We also find interesting inequalities for the values of $z$ and $\theta$ for a wide class of models with holographic
duals, generalizing analogous relations obtained earlier \cite{kiritsis,kiritsis2,trivedi,tadashi}.
The considerations in Section~\ref{sec:shape} show that
the `area law' of the entanglement entropy in the IR (modulo logarithmic corrections) requires
\beq
\theta \leq d-1, \label{thetain}
\eeq
and so we expect this inequality to apply
to generic local quantum field theories on the boundary.
Note that the UV 
always makes a contribution that obeys the area law.  Relativistic conformal field theories, with $\theta = 0$,
do obey (\ref{thetain}). However, there are supersymmetric lattice models \cite{susy} which do have large degeneracies in their spectrum, and these
could violate (\ref{thetain}).
In Section~\ref{sec:EMD}, we find that the dynamic critical exponent obeys the inequality
\beq
z \geq 1 + \frac{\theta}{d} . \label{zd}
\eeq
This inequality appears to be tied to the existence of a holographic gravity dual.
It is quite remarkable that both inequalities (\ref{thetain}) and (\ref{zd}) are realized as equalities \cite{tadashi} only for gauge theories
of non-Fermi liquid states in $d=2$, which have the exponents $\theta=1$, $z=3/2$ in
analyses to three loops \cite{metnem,metzner}. This is the first quantitative connection between the field-theoretic and holographic approaches
to non-Fermi liquids. These results also suggest that there is a strongly coupled non-Fermi liquid state in $d=3$ with
$z=5/3$, in contrast to the weak-coupling field theory analysis \cite{sonqcd} which only has marginal corrections to Fermi liquid theory.
We summarize our comparison between the field-theoretic and holographic approaches to non-Fermi liquids in Table~\ref{table}.
\begin{table}[tbp]
  \centering
   \begin{tabular}{p{7cm} c | c p{7cm}}
   Field theory &~&~& Holography \\
   \noalign{\smallskip}\hline\noalign{\smallskip}\noalign{\smallskip}\hline\noalign{\smallskip}
  A gauge-dependent Fermi surface of overdamped gapless fermions.
&$\quad$&$\quad$& Fermi surface is hidden.\\ \noalign{\smallskip}\hline\noalign{\smallskip}
  Thermal entropy density $S \sim T^{1/z}$ in $d=2$, where $z$ is the dynamic critical exponent.
&~&~&  Thermal entropy density $S \sim T^{1/z}$ in all $d$ for hyperscaling violation exponent
$\theta=d-1$, and $z$ the dynamic critical exponent. \\ 
\noalign{\smallskip}\hline\noalign{\smallskip}
  Logarithmic violation of area law of entanglement entropy, with prefactor proportional to the product of 
  $\mathcal{Q}^{(d-1)/d}$ and the boundary area of the entangling region. &~&~&
  Logarithmic violation of area law of entanglement entropy for $\theta=d-1$, with prefactor proportional to the product of 
  $\mathcal{Q}^{(d-1)/d}$ and the boundary area of the entangling region.\\ 
  \noalign{\smallskip}\hline\noalign{\smallskip}
Three-loop analysis shows $z=3/2$ in $d=2$. &~&~&
   Existence of gravity dual implies $z\geq 1+\theta/d$; leads to $z\geq 3/2$ for $\theta=d-1$ in $d=2$.\\ 
   \noalign{\smallskip}\hline\noalign{\smallskip}
Fermi surface encloses a volume proportional to $\mathcal{Q}$, as demanded by the Luttinger relation. &~&~&
The value of $k_F$ obtained from the entanglement entropy implies the Fermi surface encloses a volume proportional to $\mathcal{Q}$, as demanded by the Luttinger relation.\\ \noalign{\smallskip}\hline\noalign{\smallskip}
Gauge neutral `mesinos' reduce the volume enclosed by Fermi surfaces of gauge-charged fermions to $\mathcal{Q} - \mathcal{Q}_{\rm mesino}$.
&~&~& Gauge neutral `mesinos' reduce the volume enclosed by hidden Fermi surfaces to $\mathcal{Q} - \mathcal{Q}_{\rm mesino}$. \\
 \noalign{\smallskip}\hline
   \end{tabular}
  \caption{Comparison of the field-theoretic and holographic approaches to non-Fermi liquids (NFL). The field-theoretic results
  are as described recently in Refs.~\cite{metnem,liza}.}
  \label{table}
\end{table}

The outline of our paper is as follows. In Section~\ref{sec:EMD} we will recast the EMD
theories \cite{gubserrocha,kachru2,kachru3,trivedi,kiritsis,kiritsis2,larus1} in a general context as realizations of the IR metric (\ref{zmetric}).
We will pay particular attention to the $\mathcal{Q}$ dependence of the solution.
Various aspects of the entanglement entropy of such theories will be studied in Section~\ref{sec:shape}. We will establish (\ref{thetain}) as
a general inequality imposed by the area law of entanglement. For the hidden Fermi surface case, $\theta = d-1$, we will compute the $T=0$ entanglement entropy for an arbitrary shape, its crossover to the thermal entropy for $T>0$, and the ``mutual information'' characterizing the entanglement of disconnected regions; all of these results will be of the same form as expected for systems with Fermi surfaces, with the specific $\mathcal{Q}$ dependence displayed in Eq.~(\ref{see}).
Section~\ref{sec:fermions} will extend the EMD theories to include gauge-neutral fermions, or `mesinos', which have visible Fermi surfaces.
We will present a fully self-consistent computation, including the back-reaction of the mesinos on the metric, using the approximation
of the `electron star' theories \cite{sean1,sean2,sean3,lizasean}. We will find that the mesinos reduce the holographic entanglement entropy in precisely the manner expected from the overall Luttinger relation on the Fermi surfaces of the hidden quarks and visible mesinos \cite{liza}.
Finally some general discussion appears in Section~\ref{sec:conc}.

\section{Einstein-Maxwell-Dilaton theory}
\label{sec:EMD}

We begin with a discussion of the basic characteristics of the EMD theory of compressible quantum states \cite{trivedi,kiritsis2}
in $d$ spatial dimensions. As reviewed in Ref.~\cite{ssarcmp}, the globally conserved U(1) charge of the compressible
state is holographically realized by a U(1) gauge field $A_\mu$. In addition, the EMD theories also include a scalar field (the `dilaton') \cite{garyreview}
which is dual to a relevant perturbation on the UV conformal field theory, and which allows access to a 
wider range of IR scaling behavior in compressible states.

So we consider the holographic Lagrangian
\beq
\mathcal{L}_{EMD} =  \frac{1}{2 \kappa^2} \Bigl( R  - 2 \left(\nabla \Phi \right)^2   - \frac{V(\Phi)}{L^2} \Bigr)  - \frac{Z(\Phi)}{4 e^2}  F_{\mu\nu}F^{\mu\nu} \label{LEMD}
\eeq
defined on a $(d+2)$-dimensional spacetime with Ricci scalar $R$, with Maxwell flux $F_{\mu\nu}$ associated with $A_\mu$, 
and a dilaton field $\Phi$ with potential $V(\Phi)$ and coupling $Z(\Phi)$.

We use co-ordinates $(t, r, x_i)$, where $t$ is the time direction, $r$ is the emergent holographic direction,
and $x_i$ ($i = 1 \ldots d$) are the flat spatial directions. We will examine solutions with metric
\beq
ds^2 = L^2 \left( - f(r) dt^2 + g(r) dr^2 + \frac{dx_i^2}{r^2} \right), \label{metric}
\eeq
only the temporal component of the gauge field non-zero
\beq
A_t = \frac{eL}{\kappa} h(r), \label{vecpot}
\eeq
and a dilaton field $\Phi (r)$ dependent only upon $r$. Under these conditions, we can work with action per unit spacetime
volume of the boundary theory
\beq
\mathcal{S}_{EMD} = \int dr \frac{L^{d+2}}{r^d} \sqrt{ f(r) g(r)}\,  \mathcal{L}_{EMD} .
\eeq

The Einstein equations for this Lagrangian are
\bea
&& -\frac{4 \kappa^2 r^d \sqrt{g(r)}}{d L^d \sqrt{f(r)}} \left( f(r) \frac{\delta \mathcal{S}_{EMD}}{\delta f(r)}
- g(r) \frac{\delta \mathcal{S}_{EMD}}{\delta g(r)} \right)  \nn
&& \quad \quad \quad \quad = \frac{f'(r)}{rf(r)}+\frac{g'(r)}{rg(r)}+\frac{4}{r^{2}}+\frac{4\Phi'(r)^{2}}{d}  =  0\nn
&& \frac{2 \kappa^2 r^2 (g(r))^{3/2}}{L^d \sqrt{f(r)}} \, \frac{\delta \mathcal{S}_{EMD}}{\delta g(r)} \label{e1} \\
&& \quad \quad \quad \quad = \frac{d}{2}\frac{f'(r)}{rf(r)}-\frac{h'(r)^{2}Z(\Phi(r))}{2f(r)}-\frac{1}{2}g(r)V(\Phi(r))-\frac{d(d-1)}{2r^{2}}+\Phi'(r)^{2}  =  0, \nonumber
\eea
while the equation of motion of the dilaton field is
\bea
&& \frac{\kappa^2 r^2 \sqrt{g(r)}}{2 L^d \sqrt{f(r)}} \, \frac{\delta \mathcal{S}_{EMD}}{\delta \Phi (r)} \label{e2} \\
&& \quad \quad =
\frac{f'(r)\Phi'(r)}{2f(r)}+\frac{h'(r)^{2}Z'(\Phi(r))}{4f(r)}-\frac{g'(r)\Phi'(r)}{2g(r)}-\frac{1}{4}g(r)V'(\Phi(r))+\Phi''(r)-
d \frac{\Phi'(r)}{r}  =  0. \nonumber
\eea
Finally, the only non-zero Maxwell equation is Gauss' Law, which yields
\beq
- \frac{\kappa^2}{ L^d}
\frac{\delta \mathcal{S}_{EMD}}{\delta h (r)} = \frac{d}{dr}\left( \frac{h'(r)Z(\Phi(r))}{r^{d}\sqrt{f(r)g(r)}}\right)  =  0.
\label{e3}
\eeq
The integration constant in (\ref{e3}) is set by the charge density on the boundary \cite{ssfl},
and so we have
\beq
- \left( \frac{L^{d-1}}{\kappa e} \right) \frac{h'(r)Z(\Phi(r))}{r^{d}\sqrt{f(r)g(r)}} = \mathcal{Q}.
\label{gauss}
\eeq
The dependence of the solutions on the charge density $\mathcal{Q}$ will be crucial to our purposes.

We now discuss the structure of the solutions in the IR limit, $r \rightarrow \infty$.
As we discussed in Section~\ref{sec:intro}, we are interested in solutions which obey (\ref{zmetric}) in this limit.
Extending the results of Ref.~\cite{trivedi} to general $d$,
we can deduce that such a solution will emerge from the equations of motion provided
we choose the large $\Phi$ behavior to obey
\beq
\begin{array}{rcl} Z(\Phi) &=& Z_0 \exp \left(  \alpha \, \Phi \right) \\
V( \Phi) &=& -V_0 \exp \left( - \beta \, \Phi \right)
\end{array} ,\quad \quad \mbox{as $\Phi \rightarrow \infty$.}
\label{restrict}
\eeq
with $\alpha,\beta >0$, and the exponents $z$ and $\theta$ are then given by
\bea
\theta &=& \frac{d^2 \beta}{\alpha + (d-1) \beta} \label{thetaalpha} \\
z &=& 1 + \frac{\theta}{d} + \frac{8 (d(d-\theta) + \theta)^2}{d^2(d-\theta) \alpha^2}.  \label{zalpha}
\eea
The inequality (\ref{zd}) is clearly obeyed by (\ref{zalpha});
this inequality can also be obtained by applying the null energy
condition discussed by Ogawa {\em et al.} \cite{tadashi}. Also, imposing the inequality (\ref{thetain}) on (\ref{thetaalpha}), we obtain
\beq
\beta \leq \frac{(d-1)}{(2d-1)} \alpha . \label{alphabeta}
\eeq
The compressible state with hidden Fermi surfaces has $\theta = d-1$, and this requires
that (\ref{alphabeta}) is realized as an equality.

Inserting (\ref{restrict}) into the equations of motion, we can scale out the explicit dependence on $r$, $\mathcal{Q}$, $V_0$
and $Z_0$, by parametrizing the solution in the following form
\bea
f(r) &=& \left[ \hat{\mathcal{Q}}^{1/d} r \right]^{-2-2d (z-1)/(d-\theta)}
V_0^{-1} (V_0 Z_0)^{- \hat{\theta}} f_0 \nn
g(r) &=& \hat{\mathcal{Q}}^{2/d} \left[ \hat{\mathcal{Q}}^{1/d} r \right]^{-2+2\theta/(d-\theta)} V_0^{-1} (V_0 Z_0)^{-\hat{\theta}} g_0 \nn
h(r) &=& \left[ \hat{\mathcal{Q}}^{1/d} r \right]^{-d  - dz/(d-\theta) } h_0 \nn
e^{\Phi (r)} &=& \left[ \hat{\mathcal{Q}}^{1/d} (r/r_0) (V_0 Z_0)^{-1/(2d)} \right]^{2d(1 + \hat{\theta})/\alpha} \label{irres}
\eea
where
\beq
\hat{\mathcal{Q}} \equiv  \mathcal{Q} \frac{\kappa e}{L^{d-1}}, \label{hatQ}
\eeq
and
\beq
\hat{\theta} \equiv \frac{\theta}{d(d-\theta)}.
\eeq
The metric (\ref{irres}) is of the form (\ref{zmetric}), as expected.

To fully specify the IR solution, we need fix the values of the prefactors
$f_0$, $g_0$, $h_0$ and $r_0$ in (\ref{irres}). An analysis of the equations of motion shows that the
large $r$ limit determines the values
of 3 of these constants, $f_0/h_0^2$, $g_0$, and $r_0$; the results take a remarkably simple form when expressed
in terms of $z$ and $\theta$:
\bea
g_0  &=& (z-1)^{-\hat{\theta}} \, (z+d-\theta-1)^{1 + \hat{\theta}} \, (z+d - \theta)  \frac{d^2}{(d-\theta)^2}  \nn
\frac{f_0}{h_0^2} &=&  (z-1)^{-2 - \hat{\theta}} \, (z + d - \theta  - 1)^{1 + \hat{\theta}} \, (z+d - \theta)  \nn
r_0^{2d} &=& (z-1) \, (z+d - \theta - 1)^{-1} \label{g0f0}
\eea
The solution for the hidden Fermi surface state is realized simply by setting $\theta = d-1$ in these
general expressions.

A key feature of our results in (\ref{irres}) and (\ref{g0f0}) is that all of the exponents, and some of the scale factors, are fully determined
by the IR limit. In particular, $g_0$ and $r_0$ are fixed by (\ref{g0f0}) and so are independent of the charge
$\mathcal{Q}$. On the other hand, $f_0$ and $h_0$ are not fixed by the IR theory, only the ratio $f_0/h_0^2$ is so fixed;
these parameters
will be set by matching to appropriate constraints arising from the underlying UV theory: see Section~\ref{sec:tf} and Ref.~\cite{lizasean}.
In general, the results of this matching can depend upon the value of $\mathcal{Q}$, and so there can be additional
$\mathcal{Q}$ dependence in our results via the values of $f_0$ and $h_0$, beyond that explicitly displayed in Eq.~(\ref{irres}).
However, this UV-induced $\mathcal{Q}$-dependence does not infect the values of $g_0$, $r_0$, and $f_0/h_0^2$, and this
will be crucial below in establishing the result (\ref{see}) for the entanglement entropy.

Gauss' law plays an especially important role in this result.  Indeed, our arguments indicate that the bulk dual of the Luttinger constraint
is effectively Gauss' law \cite{ssfl,lizasean}.  As far as the solutions are concerned, Gauss' law communicates a certain special combination of UV data to the IR in a way that depends only on the intervening visible bulk charge.  The result is that the entanglement structure in the IR is sensitive to the UV only through the total charge density less the visible fermion charge, as we will see in more detail in Section~\ref{sec:fermions}.

\subsection{Non-zero temperatures}
\label{sec:temp}

The equations of motion (\ref{e1}-\ref{e3}) actually have a broader class of black hole solutions,
associated with raising the compressible state to non-zero temperatures. These solutions modify the functions $f(r)$
and $g(r)$ in (\ref{irres}) to
\bea
f_T (r) &=& f(r) \left( 1 - \left( {r}/{r_h} \right)^{d(1+z/(d-\theta))} \right) \nn
g_T (r) &=& g(r) \left( 1 - \left( {r}/{r_h} \right)^{d(1+z/(d-\theta))} \right)^{-1}
\eea
where $r_h$ is the position of the black hole horizon; the functions $\Phi (r)$ and $h(r)$ remain unchanged.

As usual, the requirement of the absence of a conical singularity at the horizon in Euclidean spacetime
fixes the temperature $T$. To determine $T$ we write $r = r_h - \rho^2$, and compute the metric in the limit
$\rho \rightarrow 0$; it has the schematic form
\beq
ds^2 = L^2 \left( - c_1 \, \rho^2 dt^2 + c_2 \, d \rho^2 + \frac{dx_i^2}{r_h^2} \right), \label{horizon}
\eeq
where $c_{1,2}$ are $\rho$-independent constants.
We now write $t = i \phi /(2 \pi T)$ and demand that the near-horizon metric have the planar contribution
$\sim \rho^2 d \phi^2 + d \rho^2$, which ensures the absence of a conical singularity provided $\phi$ is periodic
with period $2 \pi$. This fixes the Hawking temperature $T = (c_1/c_2)^{1/2} / (2 \pi)$. Such a computation yields
the following relationship between $r_h$ and $T$:
\beq
T r_h^{dz/(d-\theta)} =\hat{\mathcal{Q}}^{-z/(d-\theta)} \frac{h_0 (z+d - \theta)}{4 \pi (z-1)} \label{rhtemp}.
\eeq
The thermal entropy density, $S$, is given by the area of the horizon:
\bea
S &=& \frac{2 \pi}{\kappa^2} \frac{L^d}{r_h^d} \nn
&=& \mathcal{Q} \, T^{(d-\theta)/z} \, \frac{2 \pi eL}{\kappa}
\left[\frac{4 \pi (z-1)}{h_0 (z+d - \theta)} \right]^{(d-\theta)/z}, \label{sthermal}
\eea
and this is of the form in (\ref{st1}).
Recall that the value of $h_0$, and hence the prefactor in $S$, is not fully determined by the present IR solution.
We have to embed the solution in a UV AdS geometry to fix $h_0$, and this will be considered later in Section~\ref{sec:fermions}.
Such a determination can lead to a value of $h_0$ which is $\mathcal{Q}$-dependent, as we noted earlier. So the $\mathcal{Q}$-dependence of the thermal entropy is not only due to the factor of $\mathcal{Q}$
displayed explicitly in (\ref{sthermal}).  This is compatible with the Fermi surface interpretation, because the dispersion of low energy excitations near the Fermi surface can have a complex $\mathcal{Q}$
dependence, and this does influence in the thermal entropy. In contrast, as will see in Section~\ref{sec:shape}, the entanglement entropy has only the explicit $\mathcal{Q}$-dependence shown in (\ref{see}).

\section{Shape dependence of the entanglement entropy}
\label{sec:shape}

This section will extend the computation of the holographic entanglement entropy
by Ogawa {\em et al.} \cite{tadashi} to a general shape of the entangling region, to general $d$, and determine
the full dependence on $\mathcal{Q}$.
We will find that the results correspond to those expected from a Fermi surface.

For this computation, we will only need the equal-time metric, which we write as
\beq
ds^2 = \frac{L^2}{r^2} \left( \hat{g}_0 r^{2\theta/(d-\theta)} dr^2 + dx_i^2 \right) \label{eqmetric}
\eeq
where we initially allow for a general $\theta$, and
\beq
\hat{g}_0 \equiv \hat{\mathcal{Q}}^{2 \hat{\theta}}\,  V_0^{-1} (V_0 Z_0)^{-\hat{\theta}} g_0 .\label{hatg}
\eeq
We emphasize that the only $\mathcal{Q}$ dependence of $\hat{g}_0$ is that explicitly displayed above,
because $g_0$ is fully determined by (\ref{g0f0}).

We consider an entangling region of an arbitrary shape. However, all our considerations will be associated with
local area elements, and so we can always represent the boundary of the entangling region by solving for
one of the spatial co-ordinates, say $x_d$, in terms of the remaining co-ordinates. So the surface is (see Fig.~\ref{fig:surface})
\beq
x_d = w(x_j) \quad, \quad j = 1, \ldots, d-1
\eeq
for some function $w$ of $d-1$ co-ordinates.
We will implicitly assume below that the indices $j,j'=1, \ldots, d-1$ below. The characteristic size of the entangling region, $R$, is assumed to
be much larger than the scale set by the density $\mathcal{Q}$, and so $R \gg k_F^{-1}$; under these conditions, the entanglement entropy can be computed
using (\ref{eqmetric}), and we don't have to include the crossover to the ultimate UV behavior near the boundary.

The prescription for the holographic entanglement entropy is \cite{rt} to compute the minimal
area of a surface which encloses this entangling region in the extended holographic space. Let us parameterize the
extended surface by (see Fig.~\ref{fig:surface})
\beq
x_d = W (r, x_j). \label{hsurface}
\eeq
Then we have to find the optimum function $W(r,x_j)$
subject to the constraint
\beq
W(0,x_j) = w(x_j).
\eeq

Let us compute the area of the general holographic surface in (\ref{hsurface}).
The induced metric on this surface is
\beq
d \sigma^2  = \frac{L^2}{r^2} \left[ \left( \hat{g}_0 r^{2\theta/(d-\theta)} + \left( \frac{\partial W}{\partial r} \right)^2 \right) dr^2
 + 2 \frac{\partial W} {\partial r} \frac{\partial W}{\partial x_j} d r d x_j  +
\left( \delta_{jj'} +  \frac{\partial W} {\partial x_j} \frac{\partial W}{\partial x_{j'}} \right) d x_j d x_{j'} \right]
\eeq
The area element on the surface is determined by the square-root of the determinant of the induced metric, which is
\beq
	dA = L^d \, \hat{g}_0^{1/2} \, \frac{dr}{r^{d-\theta/(d-\theta)}} \,   d^{d-1} x_j
\left[  1 +  \left( \frac{\partial W}{\partial x_j} \right)^2  +  \frac{r^{-2\theta/(d-\theta)}}{\hat{g}_0} \left( \frac{\partial W}{\partial r} \right)^2  \right]^{1/2}
 \label{area}
\eeq
We now observe that for $d- \theta/(d-\theta) \geq 1$, which is equivalent to (\ref{thetain}),
the $r$ integral is divergent as $r\rightarrow 0$: then the leading term to the integral over $dA$
is an ultraviolet contribution proportional to $\Sigma$ (see Fig.~\ref{fig:surface}) which yields the `area law' of entanglement entropy.
Thus we expect that the inequality (\ref{thetain}) applies to holographic duals of all generic local quantum field theories which
do not have large accidental degeneracies in their low energy spectrum. Also, as we noted earlier,
relativistic conformal field theories have $\theta =0$.

The remainder of this section
limits consideration to the case $\theta = d-1$ of interest in this paper, where we have a
logarithmic violation of the area law.
Let us study the nature of the $r \rightarrow 0$ limit more carefully. Let us expand $W$ in this limit as
\beq
W(r, x_j) = w(x_j) + r^n \sigma(x_j) + \ldots \quad, \quad r \rightarrow 0,
\eeq
where it remains to determine the exponent, $n$, of the leading correction, and $\sigma$ is an arbitrary function
of the $d-1$ co-ordinates.
Inserting this in (\ref{area}) we have
\beq
dA = L^d \, \hat{g}_0^{1/2} \, \frac{dr}{r} \,  d^{d-1} x_j
\left[  1 +  \left( \frac{\partial w}{\partial x_j} \right)^2  + 2 r^n \frac{\partial w}{\partial x_j} \frac{\partial \sigma}{\partial x_j}
+  \frac{r^{2(n-d)}}{\hat{g}_0} n^2 \sigma^2 + \ldots  \right]^{1/2} \label{A1}
\eeq
The variational derivative of the integral of this expression with respect to $\sigma (x_j) $ must vanish.
A non-trivial solution is only possible if the two leading terms in powers of $r$ can cancel against each other.
So we must have $n = 2 (n-d)$ or
\beq
n = 2d. \label{nd}
\eeq
So the $r$- and $\sigma$-dependent terms inside the square-root in are indeed subdominant, and to leading logarithmic accuracy we can write
\beq
S_{E} = \frac{2 \pi}{\kappa^2} \int dA = \frac{2 \pi L^d}{\kappa^2} \, \hat{g}_0^{1/2} \, \Sigma \, \int_{r_{\text{min}}}^{r_{\text{max}}} \frac{dr}{r}
\label{A2}
\eeq
where
\beq
\Sigma = \int  d^{d-1} x_j
\left[  1 +  \left( \frac{\partial w}{\partial x_j} \right)^2 \right]^{1/2}.
\label{Sigma}
\eeq
The quantity $\Sigma$ depends only on the entangling region on the boundary, and indeed it is just its surface area.
So we conclude that the log-divergent entanglement entropy is proportional to the surface area of the entangling region,
and is otherwise independent of its shape. This is precisely the property of the entanglement entropy of a spherical
Fermi surface \cite{klich,swingle}:
our holographic analysis is for spatially isotropic systems, so a spherical Fermi surface is expected.
Also note from (\ref{hatg}) that the prefactor of (\ref{A2}) is of order $\mathcal{Q}^{(d-1)/d}$, and so the complete
$\mathcal{Q}$-dependence of the entanglement entropy is that displayed in
(\ref{see}).

Ogawa {\em et al.} \cite{tadashi} presented computations of the entanglement entropy in $d=2$ for two choices of the entangling
region: a strip and a disk. Their result for the strip agrees with our general result (\ref{A2}), and also with the
exponent value in (\ref{nd}). They presented numerical results for the disk, and the prefactor of their logarithm equals that predicted
by (\ref{A2}).

We also wish to point out that this result may be interpreted as additional evidence for the Ryu-Takayanagi formula.  While their formula has been proven for spheres \cite{myers}, it remains unproven in general and is known to be modified in higher derivative gravity.  In general, we can only show that the Ryu-Takayanagi proposal has the right basic structure to give an entanglement entropy.  Our calculation here shows that the Ryu-Takayanagi formula reproduces in detail a universal feature of the entanglement structure of compressible states as expected from field theory for all region shapes.

Finally, let us discuss the limits on the $r$ integration in (\ref{A2}). We expect the large $r$ limit to be set by the size of the entangling
region, $R$. From (\ref{scaling}) and (\ref{rscaling}), we see that for $\theta =d-1$, $r^d$ scales as $x_j \sim R$. So we can expect $k_F r_{\text{max}} \sim (k_F R)^{1/d}$.
We will discuss the value of $r_{\text{min}}$ more carefully in Section~\ref{sec:fermions}, where we will argue that
$k_F r_{\text{min}} \sim 1$. Note that $r_{\rm max} \gg r_{\rm min}$ because $k_F R \gg 1$.
With these limits, and using $\mathcal{Q} \sim k_F^d$ and $\Sigma \sim R^{d-1}$,
we see that (\ref{A2}) is of the form (\ref{see}).

\subsection{Entanglement to thermal crossover}
Using our finite temperature solution, we may also study the von Neumann entropy of a subsystem at finite temperature.  There will be a universal crossover function that connects the anomalous entanglement entropy to the thermal entropy discussed above.  This crossover function is universal because the same low energy degrees of freedom responsible for the long range entanglement also control the low temperature thermodynamics.  However, this universality is only defined up to boundary law terms since these may be generated by UV processes and can depend on the UV cutoff.  On general scaling grounds, this crossover function will depend on region size $R$ and temperature $T$ according to
\beq
S(R,T) = \mathcal{Q}^{(d-1)/d} T^{\phi} F_S(R T^{1/z}).
\eeq
Note that we have suppressed a dimensionful constant multiplying $RT^{1/z}$ from the dispersion relation analogous to the Fermi velocity of a Fermi liquid.  The scaling function itself is independent of this constant.  A general discussion of such scaling forms will appear very soon in \cite{swingle2} where it is shown that $\phi = (1-d)/z$ for a compact region.  This result is obtained by matching onto the thermodynamic entropy where extensivity requires $F_S(x) \rightarrow x^d$.  We may obtain this function holographically by computing areas of bulk minimal surfaces in black hole backgrounds.  We set $\theta = d-1$ in this section.

We consider a strip-like entangling region of cross section $\Sigma/2$ and width $R$.  Assuming $R^{d-1} \ll \Sigma$ then the total surface area is $\Sigma$.  Because of the translation invariance in the cross-sectional directions, the bulk coordinate $r$ will be a function of only one boundary coordinate $r=r(x)$.  The area is
\beq
A = \left(\frac{\Sigma}{2}\right) \int^{R/2}_{-R/2} dx \left(\frac{L}{r}\right)^d \sqrt{1 + \hat{g}_0 r^{2(d-1)}\left(1 - \left(r/r_h\right)^{d(1+z)}\right)^{-1} \left( \frac{dr}{dx}\right)^2},
\eeq
and the resulting equation of motion has an integral of the motion given by
\beq
\frac{1}{r^d \sqrt{1 + \hat{g}_0 r^{2(d-1)} \left(1 - \left(r/r_h\right)^{d(1+z)}\right)^{-1} (dr/dx)^2}} = \frac{1}{r_{\text{max}}^d}.
\eeq
Solving for $dr/dx$ we find
\beq
\frac{dr}{dx} = \frac{\sqrt{r_{\text{max}}^{2d}-r^{2d}}}{r^{2d-1} \sqrt{\hat{g}_0 \left(1 - \left(r/r_h\right)^{d(1+z)}\right)^{-1}} }.
\eeq

We now recast the area as an integral over $r$ to obtain
\beq A = \left(\frac{\Sigma}{2}\right) 2 \int_{r_{\text{min}}}^{r_{\text{max}}} dr \frac{r^{2d-1} \sqrt{\hat{g}_0}}{\sqrt{r_{\text{max}}^{2d}-r^{2d}}\sqrt{1 - \left(r/r_h\right)^{d(1+z)}}} \frac{L^d r_{\text{max}}^{d}}{r^{2d}} \label{stripA}
\eeq
and
\beq
R = 2 \int_{r_{\text{min}}}^{r_{\text{max}}} dr \frac{r^{2d-1} \sqrt{\hat{g}_0}}{\sqrt{r_{\text{max}}^{2d}-r^{2d}}\sqrt{1 - \left(r/r_h\right)^{d(1+z)}}}. \label{stripR}
\eeq
We may safely set the UV cutoff $r_{\text{min}}=0$ in the integral for $R$.  This gives a universal relationship between $R$, $r_{\text{max}}$, and $r_h$.  Changing variables to $u = {r}/{r_{\text{max}}}$ we find
\beq
R = 2 \sqrt{\hat{g}_0} r_{\text{max}}^d  \int^1_0 du \frac{u^{2d-1} }{\sqrt{1-u^{2d}}\sqrt{1 - \left(r_{\text{max}}/r_h\right)^{d(1+z)} u^{d(1+z)}}}.\eeq
According to (2.22) we have $r_h^d \sim T^{-1/z}$ and upon putting $2 \sqrt{\hat{g}_0} r^d_{\text{max}} = R \,G(R T^{1/z})$ we find
\beq
1 = G(R T^{1/z}) \int^1_0 du \frac{u^{2d-1} }{\sqrt{1-u^{2d}}\sqrt{1 - u^{d(1+z)} \left( R T^{1/z} \,G(R T^{1/z})\right)^{1+z}}}
\eeq
where we have absorbed a dimensionful constant into $RT^{1/z}$.  This constant can depend on $\mathcal{Q}$ as well as other UV data (through $h_0$ as in (\ref{rhtemp})) just as the Fermi velocity in a Fermi liquid can, but as we have shown explicitly here, the form of the scaling function is independent of this constant.

We now determine the function $G$ in certain limits.  Starting from
\beq
1 = G(x) \int^1_0 du \frac{u^{2d-1} }{\sqrt{1-u^{2d}}\sqrt{1 - u^{d(1+z)} (x G(x))^{1+z}}}
\eeq
we see immediately that  $G(x)$ goes to a constant as $x\rightarrow 0$.  On the other hand, as $x\rightarrow \infty$ we have $G(x) = 1/x$ to keep the integral well defined.  This immediately gives $r^d_{\text{max}} \sim r^d_h$ as expected.  To compute the subleading term at large $x$ we define $G(x) = (1-\delta)/x$ which gives
\beq
1 = \frac{1}{x} \int^1_0 du \frac{u^{2d-1} }{\sqrt{1-u^{2d}}\sqrt{1 - u^{d(1+z)} (1-(z+1)\delta)}}.
\eeq
There is a logarithmic singularity as $\delta \rightarrow 0$ coming from $u \rightarrow 1$, so to extract it we set $u = 1-v$ and expand
\beq
x = \int^{\mathcal{O}(1)}_0 dv \frac{1}{\sqrt{2d v}\sqrt{d(1+z)v + (1+z)\delta}}.
\eeq
This gives
\beq
x \sqrt{2 d^2 (1+z)} \sim \ln{(1/\delta)}
\eeq
and hence $\delta \sim \exp{(- d \sqrt{2(1+z)} x)}$ with $x = RT^{1/z}$.

Having established a scaling form for $r_{\text{max}}$, a similar form immediately follows for the area itself.  We have
\beq
A = \left(\frac{\Sigma}{2}\right) 2 \sqrt{\hat{g}_0} L^d \int^1_{r_{\text{min}}/r_{\text{max}}} \frac{du}{u} \frac{1}{\sqrt{1-u^{2d}}\sqrt{1 - u^{d(1+z)} (RT^{1/z} G(RT^{1/z}))^{1+z}}}.
\eeq
Using $\sqrt{\hat{g}_0} \sim \mathcal{Q}^{(d-1)/d}$ we see that the area reduces to
\beq
\mathcal{Q}^{(d-1)/d} \Sigma \ln{(\mathcal{Q}^{1/d} R)}
\eeq
for $RT^{1/z}\rightarrow 0$ and to
\beq
\mathcal{Q}^{(d-1)/d} \Sigma R T^{1/z}
\eeq
for $RT^{1/z}\rightarrow \infty$.

The crossover function has $\phi=0$ instead of $\phi = (1-d)/z$ because we worked in a limit where the cross-section $\Sigma$ was larger than any other scale.  In that limit matching to the thermodynamic result demands $F_S(x) \rightarrow x$ instead of $F_S(x) \rightarrow x^d$ and hence $\phi =0$.

The interpretation of this result in the field theory is simple: we have a $d-1$ dimensional manifold (the Fermi surface) of gapless modes propagating in one effective (radial) direction with dynamic exponent $z$.  At finite temperature these modes give an entropy of the form $\mathcal{Q}^{(d-1)/d} \Sigma R T^{1/z}$ while at zero temperature each mode gives a $\ln{(\mathcal{Q}^{1/d}R)}$ contribution to the entanglement entropy for a total entropy of $\mathcal{Q}^{(d-1)/d} \Sigma \ln{(\mathcal{Q}^{1/d} R)}$.

\subsection{Mutual information}
We can also compute the entanglement entropy for disjoint regions using the Ryu-Takayanagi formula \cite{rt}.  Such configurations are interesting in part because they can be used to define subtracted versions of the entanglement entropy that are not UV sensitive.  For two distant regions we find that $S(AB) = S(A) + S(B)$.  As the regions are brought close, the bulk minimal surface experiences a phase transition and suddenly connects the two regions in the bulk.  Using $S(A)$, $S(B)$, and $S(AB)$ we may compute the mutual information $I(A,B) = S(A) + S(B) - S(AB)$.  This quantity is guaranteed to be positive and has a variety of other favorable properties.  It is also insenstive to the UV and measures both quantum and classical correlations.

\begin{figure}[tbp]
  \centering
  \includegraphics[width=4in]{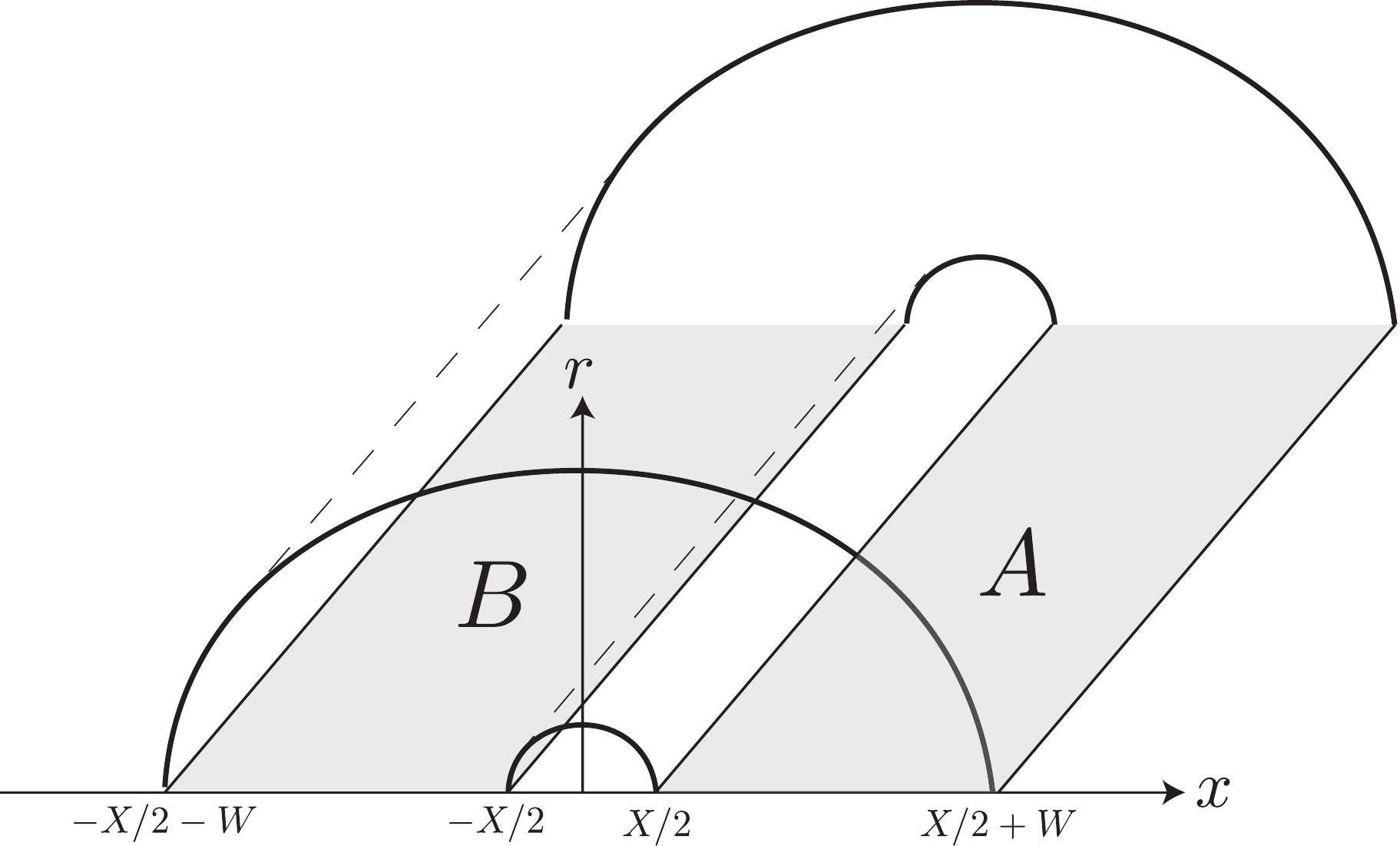}
  \caption{Geometry of mutual entanglement bewteen regions $A$ and $B$.
  }
  \label{fig:mutual}
\end{figure}
For two strips of cross-section $\Sigma/2$, width $W$, and separation $X$, and with $\mathcal{Q}^{-1/d} \ll X \ll W$
(see Fig.~\ref{fig:mutual}),
we computed the holographic entanglement entropy from (\ref{zmetric}) for $\theta = d-1$, and
found a mutual information of the form
\beq
I \sim \mathcal{Q}^{(d-1)/d} \Sigma \ln{(W/X)} \label{mI}
\eeq
which is identical to the form obtained from a Fermi surface \cite{swingleMI}.  For some critical value of $X$ of order $W$, the mutual information will drop continuously to zero and remain there as $X$ is increased.  In principle, the long distance decay of the mutual information should contain information about the decay of correlation functions, but the geometric prescription is blind to this subleading contribution.

The geometric computation proceeds from (\ref{stripA}) and (\ref{stripR}) which give the area for a minimal surface of boundary width $R$.  We set $r_h=\infty$ in this section.  Region $A$ is a strip between $x=X/2$ and $x=X/2+W$ and region $B$ is a strip between $x=-X/2-W$ and $x=-X/2$.  The entropies $S(A)$ and $S(B)$ are identical and proportional to (\ref{stripA}) with $R = W$.  For $X \ll W$, the entropy $S(AB)$ consists of two pieces, one associated with a surface connecting $x=-X/2-W$ to $x=X/2+W$ and the other associated with a surface connecting $x=-X/2$ to $x=X/2$ (see Fig.~\ref{fig:mutual}).  These surfaces both obey (\ref{stripA}) and (\ref{stripR}) but with $R=X+2W$ and $R=X$ respectively.

Setting $r_h = \infty$ in (\ref{stripA}) and (\ref{stripR}) and changing variables to $r = r_{\text{max}} u$ we have
\beq
A = \left(\frac{\Sigma}{2}\right) (2 L^d \sqrt{\hat{g}_0}) \int^1_{r_{\text{min}}/r_{\text{max}}} \frac{du}{u} \frac{1}{\sqrt{1-u^{2d}}}
\eeq
and
\beq
R = (2 \sqrt{\hat{g}_0} r^d_{\text{max}} ) \int^1_{r_{\text{min}}/r_{\text{max}}} \frac{du}{u} \frac{u^{2d}}{\sqrt{1-u^{2d}}}.
\eeq
Taking $r_{\text{min}} = \epsilon \rightarrow 0$ we find
\beq
A = \left(\frac{\Sigma}{2}\right) (2 L^d \sqrt{\hat{g}_0}) \ln{\left(\frac{r_{\text{max}}}{\epsilon}\right)}
\eeq
and
\beq
R = 2 \sqrt{\hat{g}_0} r^d_{\text{max}} k
\eeq
with $k = \int_0^1 du u^{2d-1}/\sqrt{1-u^{2d}}$.  Combining these two equations immediately gives
\beq
A(R) =\left(\frac{\Sigma}{2}\right) (2 L^d \sqrt{\hat{g}_0}) \frac{1}{d} \ln{\left(\frac{R}{2k \sqrt{\hat{g}_0} \epsilon^d}\right)}. \label{stripAR}
\eeq
The mutual information is
\beq
I \propto A(W) + A(W) - (A(X) + A(X+2W))
\eeq
which gives
\beq
\left(\frac{\Sigma}{2}\right) (2 L^d \sqrt{\hat{g}_0}) \frac{1}{d} \ln{\left(\frac{W^2}{X(X+2W)}\right)}.
\eeq
Note that the UV cutoff $\epsilon$ has vanished and that in the limit $X \ll W$ the argument of the log simplifies to $W/X$ as claimed
in (\ref{mI}).

\section{Adding mesinos}
\label{sec:fermions}

All the analysis of this section will be restricted to spatial dimension $d=2$, and $\theta =d-1$.

This section will add fermionic matter to the NFL phase of the EMD theory, in a manner analogous to previous computations \cite{sslee0,hong0,zaanen1,hong1,denef,faulkner,polchinski,gubserrocha,hong2,sean1,sean2,sean3,larus2,trivedi,leiden,hong4,yarom,ssfl,larus3,lizasean}. These fermions represent gauge neutral `mesinos' on the boundary, in the language of Refs.~\cite{ssfl,lizasean}.
When a non-zero density of such mesinos is present, we are in a FL* phase, and there is a visible Fermi surface, which encloses a volume associated with the mesino
density, $\mathcal{Q}_{\rm mesino}$. We are primarily interested here in the influence of this visible Fermi surface
on the hidden Fermi surfaces.

The mesinos occupy fermionic states
obeying the Dirac equation which can be written
in terms of a two-component spinor, $\chi_k$ \cite{hong1,trivedi,ssfl} :
\beq
\left( - i \sigma^y  \sqrt{\frac{f(r)}{g(r)}} \frac{d}{dr} - mL \sigma^x \sqrt{f(r)}  - k \sigma^z r \sqrt{f(r)}  - q A_t (r) \right) \chi_{k} (r) = E \, \chi_{k} (r)
\label{dirac}
\eeq
where $q$ is the conserved $\mathcal{Q}$-charge of the mesinos,
$\sigma$ are Pauli matrices acting on the spinor space, $k$ is the momentum of the state along the boundary,
and $E$ is the energy eigenvalue. The mass $m$ determines the scaling dimension of the mesonic operator
on the UV conformal field theory, and we
will impose the UV boundary condition  $\chi ( r \rightarrow 0) \sim r^{mL}$,
and outgoing wave restrictions for $r \rightarrow \infty$.
The action associated with these fermionic states can be obtained by evaluating the
determinant of the Dirac operator, as discussed in Ref.~\cite{denef}, which is expressed as a sum
over quasi-normal modes.
This is a rather involved computation, but we can obtain some features of interest to us by
using the structure of the solutions of (\ref{dirac}), which were discussed in some detail by Iizuka
{\em et al.} \cite{trivedi}. For the metric (\ref{zmetric}), we are in the ``Fermi liquid'' regime examined
by Iizuka {\em et al.}: the quasi-bound states are well-defined near the Fermi energy, $E \rightarrow 0$, with
an exponentially small decay rate which can be computed by WKB methods.
For now, let us ignore the tunneling decay process. Then we have discrete set of energy eigenstates $E_\ell (k)$, labeled by the discrete
index $\ell$, just as in the simpler situation discussed in Ref.~\cite{ssfl}. The action of these fermionic states is obtained
by adding up the energy of all the occupied negative energy states \cite{denef,ssfl}
\beq
\mathcal{S}_{\rm mesino} = \sum_\ell \int \frac{d^2 k}{4 \pi^2} \, E_\ell (k) \, \theta \left(- E_\ell (k) \right)
\label{sf1}
\eeq
If we included the decay processes, this expression would be replaced by the determinant of the Dirac
operator \cite{denef}.

We are now in a position to write down the corrections to the equations of motion (\ref{e1}), (\ref{e2}), (\ref{e3})
due to the presence of the mesinos. The equations now follow from the stationarity conditions on the complete
action $\mathcal{S}_{EMD} + \mathcal{S}_{\rm mesino}$. The functional derivatives of
$\mathcal{S}_{EMD}$ were given in (\ref{e1}), (\ref{e2}), (\ref{e3}), while those of
$\mathcal{S}_{\rm mesino}$
can be evaluated using the Feynman-Hellman theorem \cite{ssfl}: we normalize the fermion states so that
\beq
\int dr \, \sqrt{\frac{g(r)}{f(r)}}  \chi_{\ell,k} ^\dagger (r) \chi_{\ell,k} (r) = 1. \label{norm}
\eeq
and then the  functional derivatives of $\mathcal{S}_{\rm mesino}$ follow from those of $E_\ell (k)$:
\bea
\frac{\delta E_\ell (k)}{\delta A_t (r)} &=& - q \, \chi_{\ell,k}^\dagger (r) \chi_{\ell,k} (r)  \sqrt{\frac{g(r)}{f(r)}}  \nn
\frac{\delta E_\ell (k)}{\delta f (r)} &=&  \sqrt{\frac{g(r)}{f(r)}} \chi_{\ell,k}^\dagger (r) \left( -  \frac{i \sigma^y}{2}  \frac{1}{\sqrt{f(r)g(r)}} \frac{d}{dr}  - \frac{m L}{2 \sqrt{f(r)}}  \sigma^x - \frac{k r}{2 \sqrt{f(r)}}  \sigma^z    \right) \chi_{\ell,k} (r)  \nn
\frac{\delta E_\ell (k)}{\delta g (r)} &=&  \chi_{\ell,k}^\dagger (r) \left(   \frac{i \sigma^y}{2}  \frac{1}{g(r)} \frac{d}{dr}\right)  \chi_{\ell,k} (r)
\label{sf2}
\eea
For our purposes, the most important modification is to Gauss' Law, which is modified from (\ref{gauss}) to
\beq
- \frac{L}{\kappa e} \frac{h'(r)Z(\Phi(r))}{r^{2}\sqrt{f(r)g(r)}} + q \sum_\ell \int \frac{d^2 k}{4 \pi^2} \,
\theta \left(- E_\ell (k) \right) \int_0^r dr' \chi_{\ell,k}^\dagger (r') \chi_{\ell,k} (r')  \sqrt{\frac{g(r')}{f(r')}} =  \mathcal{Q}.
\label{gaussf}
\eeq

These equations are too difficult to solve in general, but we can easily identify the characteristic
scale of $r$ at which the mesino states are present. And this follows by generalizing the dimensional
arguments leading to (\ref{irres}). Using the $\mathcal{Q}$ dependence shown explicitly in (\ref{irres}) ,
we deduce that the $\mathcal{Q}$-dependence of the Dirac wavefunction can be written in the scaling form
\beq
\chi_k (r) = F_k ( \mathcal{Q} r^2) \label{chiscale}
\eeq
where we reiterate that we are in $d=2$. Also, recall that below (\ref{irres}) we discussed additional $\mathcal{Q}$-dependence, arising from the UV matching,
in the values of $f_0$ and $h_0$; examination of the structure of (\ref{dirac}), while using the fact that $f_0/h_0^2$ is $\mathcal{Q}$-independent, leads to the conclusion that this additional $\mathcal{Q}$-dependence does {\em not\/}
infect the scaling form in
(\ref{chiscale}). It does, however, appear in the value of the eigenvalue $E$.
From these results we conclude that all the occupied mesino states are at
$r \sim \sqrt{\mathcal{Q}}$.

A similar conclusion is reached by a WKB analysis of the Dirac equation, along the lines of Ref.~\cite{sean3}.
By an appropriate change of variables, the Dirac equation is converted to a Schr\"odinger equation
for the top component of the wavefunction, and the mesino wavefunction is then concentrated in the classically
allowed region of the potential: this argument also leads to a characteristic $r \sim \sqrt{\mathcal{Q}}$.

We now look at the regime $r \gg \sqrt{\mathcal{Q}}$ which is relevant for the computation of the
entanglement entropy. The mesino states are at a scale $r \sim \sqrt{\mathcal{Q}}$,
and we will ignore their possible contribution from tunneling out of the classically allowed region
into $r \gg \sqrt{\mathcal{Q}}$. The possibility remains that these tunneling processes have a strong
back-reaction on the metric, but we will not study this here. With this assumption, we examine
Gauss' Law for $r \gg \sqrt{\mathcal{Q}}$: we can write (\ref{gaussf}) as \cite{lizasean}
\beq
- \frac{L}{\kappa e} \frac{h'(r)Z(\Phi(r))}{r^{2}\sqrt{f(r)g(r)}} + \mathcal{Q}_{\rm mesino} =  \mathcal{Q},
\quad\quad \quad r \gg \sqrt{\mathcal{Q}}.
\label{gaussfls}
\eeq
where, after using (\ref{norm}), we have
\beq
\mathcal{Q}_{\rm mesino} =  q \sum_\ell \int \frac{d^2 k}{4 \pi^2} \,
\theta \left(- E_\ell (k) \right) \label{qmesino}
\eeq
is the total charge density in the occupied mesino states.
So for $r \gg \sqrt{\mathcal{Q}}$, (\ref{gaussfls}) replaces (\ref{gauss}), while the differential Einstein equations
(\ref{e1}) remain unchanged.
Also, there is no mesino contribution to the dilaton equation (\ref{e2}), which remains unchanged.
So the solutions for $r \gg \sqrt{ \mathcal{Q}}$ are as in (\ref{irres}) and (\ref{g0f0}), but with
$\mathcal{Q} \rightarrow \mathcal{Q} - \mathcal{Q}_{\rm mesino}$ in the explicitly displayed
$\mathcal{Q}$ dependence. This is the key result of this section so far.

\begin{figure}[tbp]
  \centering
  \includegraphics[width=3.8in]{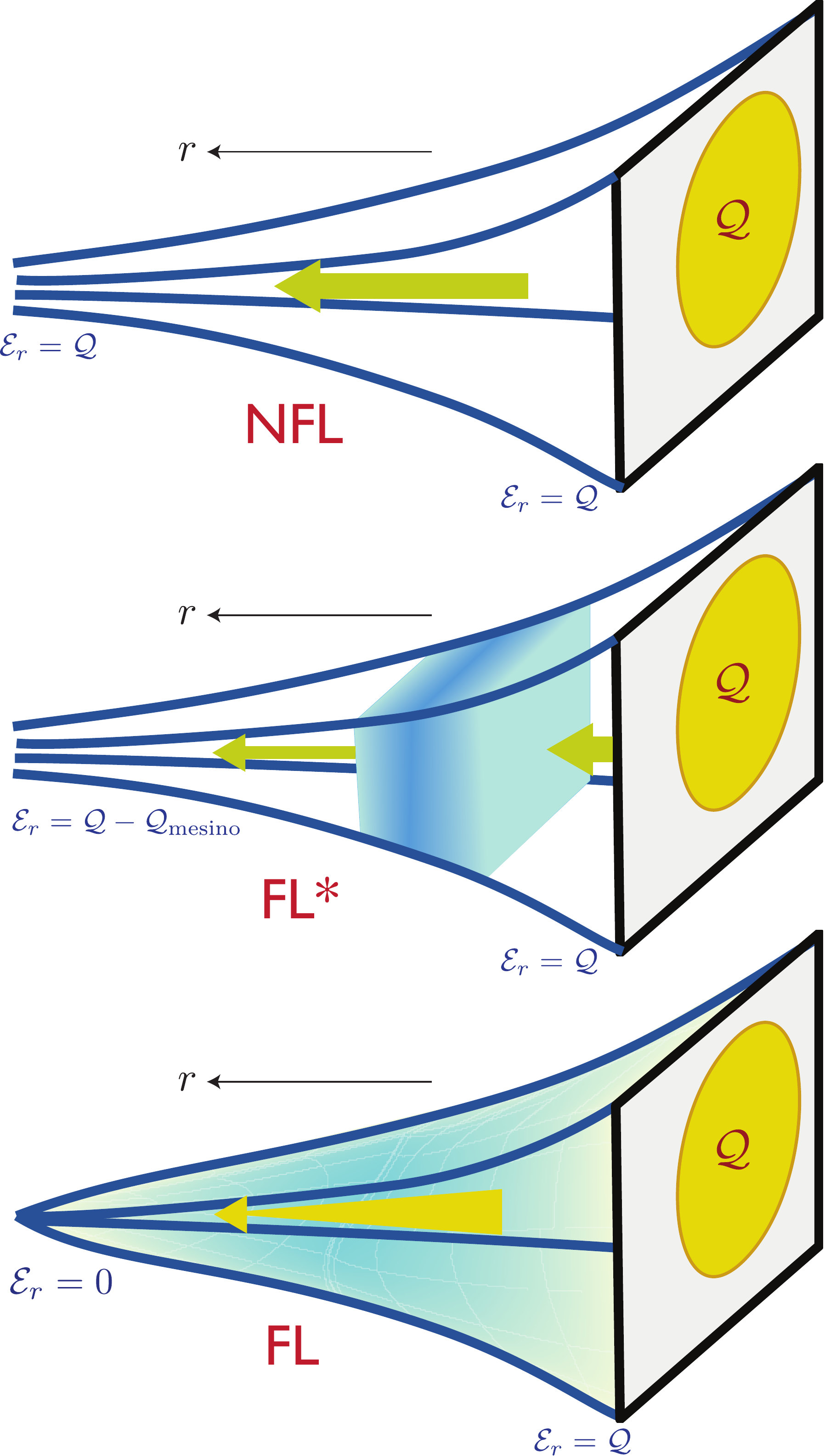}
  \caption{Schematic illustration of the holographic geometries of various compressible phases. The phases are labeled non-Fermi liquid (NFL), fractionalized Fermi liquid (FL*), and Fermi liquid (FL), following the notation of Ref.~\cite{liza}. The boundary theory
  at $r=0$ has total charge density $\mathcal{Q}$
  which sources the bulk electric field $\mathcal{E}_r$; we define $\mathcal{E}_r$ to equal the
  left-hand-side of (\ref{gauss}). The NFL phase is described by the theory in Section~\ref{sec:EMD}, and has the 
  IR metric (\ref{zmetric}). The shading in the bulk region represents the density of mesinos. The FL* phase is described in Section~\ref{sec:tf} using the Thomas-Fermi approximation. The FL* phase has
  mesinos are in the bulk shaded region at $r \sim \sqrt{\mathcal{Q}}$, which
  this corresponds to the shaded region in Fig.~\ref{fig:fluid}; the region $r \gg \sqrt{\mathcal{Q}}$, with 
  $\mathcal{E}_r = \mathcal{Q} - \mathcal{Q}_{\rm mesino}$, determines the entanglement entropy. The FL phase is not
  described in the present paper: its geometry is confining and terminates at a finite $r$ where $\mathcal{E}_r =0$, as discussed in Ref.~\cite{ssfl}.
  }
  \label{fig:ffl}
\end{figure}

Also note that (\ref{e1}), (\ref{e2}) and (\ref{gaussfls}) do not determine the values of $f_0$ and
$h_0$. So, in general, the presence
of the mesinos will change the values of $f_0$ and $h_0$,
beyond that associated with $\mathcal{Q} \rightarrow \mathcal{Q} - \mathcal{Q}_{\rm mesino}$. However, $g_0$ was fully determined by (\ref{e1}), (\ref{e2}) and (\ref{gaussfls}), and this is the only parameter
that appears in the holographic entanglement entropy. So we can simply apply the computation of the holographic entanglement entropy
of Section~\ref{sec:shape} after  the substitution $\mathcal{Q} \rightarrow \mathcal{Q} - \mathcal{Q}_{\rm mesino}$.
The present argument also shows that this computation applies for $r \gg \sqrt{\mathcal{Q}}$, and
so $r_{\rm min} \sim \sqrt{\mathcal{Q}}$. We schematically illustrate this result in Fig.~\ref{fig:ffl}.

So, as claimed earlier, we have established that the main effect of the occupied mesino states in the FL* phase
is to subtract away from the charge appearing in the holographic entanglement entropy. This is consistent
with the expected transfer of quarks to mesino states, and the corresponding expected decrease in the hidden Fermi surface
volume of the quark states. This feature provides additional support to the hidden Fermi surface interpretation of the NFL state.

\subsection{Thomas-Fermi theory}
\label{sec:tf}

Our analysis of the Dirac equation above treats the mesino states in a Hartree-Fock-like theory.
The resulting equations are quite complicated to solve, and so we will consider here the analog
of the simpler Thomas-Fermi theory. In the present context, the Thomas-Fermi theory is the `electron star' approach \cite{sean1,sean2,sean3,lizasean},
in which the electrons are treated as a continuous
fluid which obeys the equation of state of free Dirac fermions in a local chemical potential.
This fluid resides in the classically allowed region of the potential associated with the
Dirac equation \cite{sean3}.
In this approach, we will be able to
embed our IR solution in an EMD theory which has a AdS$_4$ metric in the UV. We will compute the transition
from the NFL state to a FL* state with occupied mesino states, while including the back-reaction of the mesinos
on the metric. We will also find here that the mesino states appear
at a characteristic $r \sim \sqrt{\mathcal{Q}}$.

In \cite{lizasean} is was shown that one can study charge fractionalization in the EMD theories by dialing a relevant coupling in the dual UV theory. This coupling is dual to the dilaton field and can be made relevant by choosing the dilaton mass appropriately. The analysis presented in this subsection will closely follow \cite{lizasean}. We consider the EMD theory (\ref{LEMD}), and add the contribution from the Dirac fluid, which is simply given by the pressure
\beq
\mathcal{L}_{\rm fluid}=p(\mu_{\rm loc}).
\eeq
Here $\mu_{\rm loc}$ stands for the local chemical potential, which for irrotational, zero temperature fluids can be taken to be
\beq
\mu_{\rm loc}=\frac{A_t}{\sqrt{-g_{tt}}}=\frac{e}{\kappa}\frac{h}{\sqrt{f}},
\eeq
where we used (\ref{metric}) and (\ref{vecpot}). The pressure of the fluid can be expressed in terms of the energy and charge density
\beq
-\hat{p}=\hat{\rho}-\frac{h}{\sqrt{f}} \hat{\sigma}, \quad \hat{\sigma}=\hat{\beta} \int_{\hat{m}}^{\frac{h}{\sqrt{f}}} \epsilon \sqrt{\epsilon^2-\hat{m}^2}d\epsilon, \quad \hat{\rho}=\hat{\beta} \int_{\hat{m}}^{\frac{h}{\sqrt{f}}} \epsilon^2 \sqrt{\epsilon^2-\hat{m}^2}d\epsilon,
\eeq
where we introduced dimensionless variables
\bea
p = \frac{1}{L^2 \kappa^2} \hat{p} \,, \qquad \rho = \frac{1}{L^2 \kappa^2} \hat{\rho} \,, \qquad \sigma = \frac{1}{e L^2 \kappa} \hat{\sigma} \,, \nn
\hat{\beta} = \frac{e^4 L^2}{\kappa^2} \frac{1}{\pi^2} \,, \qquad \hat{m^2} = \frac{\kappa^2}{e^2} m^2 \,.
\eea
The equations of motion for the Einstein-Maxwell-dilaton-charged fluid theory for the metric and vector potential ansatzes (\ref{metric}) and (\ref{vecpot}) are found to be
\bea
\frac{1}{r} \left(\frac{f'}{f} + \frac{g'}{g} + \frac{4}{r} \right) + \frac{g h}{\sqrt{f}} \, \hat{\sigma} + 2 \Phi'^2 & = & 0 \,,  \nn
\frac{1}{r} \left(\frac{f'}{f} - \frac{1}{r} \right) + g \left( \hat{p} - \frac{1}{2} V(\Phi) \right) - \frac{Z(\Phi) h'^2}{2 f} + \Phi'^2 & = & 0 \,, \nn
- \Phi'' + \frac{1}{2} \left(- \frac{f'}{f} + \frac{g'}{g} + \frac{4}{r} \right) \Phi' + \frac{g V'(\Phi)}{4} - \frac{Z'(\Phi) h'^2}{4 f} & = & 0 \,,  \nn
\frac{d}{dr} \left(\frac{Z(\Phi) h'}{r^2 \sqrt{f g}}\right) - \frac{\sqrt{g}}{r^2} \hat{\sigma} & = & 0 \,. \label{eq:EMDF}
\eea
One easily verifies that the above equations reduce to the EMD equations of motion (see Section \ref{sec:EMD}) in the absence of the fluid, that is when $\hat{\mu}_{\rm loc}<\hat{m}$. As in the previous section we see that the presence of the fermions in the bulk modifies Gauss' Law. In the fluid approximation the total charge carried by the fermions is
\beq
\hat{\mathcal{Q}}_{\rm mesino}= \int_{r_1}^{r_2} \frac{\sqrt{g}}{r^2}\hat{\sigma},
\eeq
where $r_1$ and $r_2>r_1$ are the radii between which a non-zero density of the fluid is present. The total charge now reads $\hat{\mathcal{Q}}=\hat{\mathcal{Q}}_{\rm mesino}+\hat{\mathcal{Q}}_{\rm quark}$, where
\beq
\hat{\mathcal{Q}}_{\rm quark}=-\left(\frac{Z(\Phi (r)) h'(r)}{r^2 \sqrt{f(r) g(r)}}\right) \quad \textrm{for $r>r_2$}.
\eeq
To explore the $\mathcal{Q}$ dependence of theory, it is useful to note here that
the equations of motion, (\ref{eq:EMDF}), are invariant under the scale transformations
\bea
r & \rightarrow & \lambda \,  r \nn
f & \rightarrow & f \nn
g & \rightarrow & \lambda^{-2} g \nn
h & \rightarrow & h \nn
\Phi & \rightarrow & \Phi \nn
\mathcal{Q}_{\rm quark} & \rightarrow & \lambda^{-d} \mathcal{Q}_{\rm quark}, \label{Qscale}
\eea
where $\lambda$ is the rescaling factor, and the additional rule that $\{\hat{\sigma}, \hat{\rho}, \hat{p}\} \to \{\hat{\sigma}, \hat{\rho}, \hat{p}\}$.

To embed the theory in an asymptotic AdS$_4$ metric, we choose the dilaton potential to take the form near $\Phi=0$ of
\beq
V(\Phi)=-6+2 M_{\Phi}^2 L^2 \Phi^2+\mathcal{O}(\Phi^4).
\eeq
For computational convenience we will take the dilaton mass to satisfy $M_{\Phi}^2=-2/L^2$, such that the dual operator $\mathcal{O}$ has scaling dimension $\Delta=2$. Note that the mass is above the Breitenlohner-Freedman bound. If we also choose $Z(0)=1$, the near boundary expansion of the dilaton field takes the form
\beq
\Phi=\phi_0 r+\frac{\langle \mathcal{O} \rangle}{2} r^2+\dots \label{phiUV}
\eeq
where $\phi_0$ and $\langle \mathcal{O} \rangle$ are the source and the expectation value of the relevant operator $\mathcal{O}$ dual to the bulk dilaton $\Phi$.

It is now convenient to fix $V(\Phi)$ and $Z(\Phi)$ such that they interpolate between the AdS$_4$ solution in the UV and the IR solution described in Section \ref{sec:EMD}. Note that indeed for the latter solution, (\ref{irres}), the local chemical potential goes to zero in the IR, which implies that the fluid is absent. We expect our conclusions to be insensitive to the specific interpolation form, and we found it convenient to use the following expressions
\bea
V(\Phi) &=& - \frac{V_0}{2 \cosh (\alpha \Phi/3)}+\left(\frac{1}{2}V_0-6 \right)(1-\tanh(\alpha \Phi/3)^2) ,\nn
Z(\Phi) &=& \exp(\alpha \Phi),
\eea
with $V_0=24 \left(\alpha ^2+6\right)/{\alpha ^2}$ to get the desired dilaton mass in the UV. One readily verifies that these expressions take the form of (\ref{restrict}) with $d=2$ in the limit of $\Phi \to \infty$ in the IR. Finally, without loss of generality we set $\alpha=3$ for computational convenience.

For the full potential, the IR solution is modified by a series expansion together with a perturbation that will allow the flow to be integrated up to the UV. Furthermore, the generalized Gauss' Law implies that we have to replace $\hat{\mathcal{Q}}$ by $\hat{\mathcal{Q}}_{\rm quark}$ in (\ref{irres}). The IR solution thus reads:
\bea
f & = & \frac{63}{10^4} \sqrt{\frac{7}{2}} \frac{h_0^2}{\hat{\mathcal{Q}}_{\rm quark}^6 r^{12}} \left(1 + \sum_{n=1}^\infty \frac{f_n}{r^{2n}} + \delta f \, r^{N} \right) \,, \label{eq:ffa} \\
g & = & \frac{63}{40} \hat{\mathcal{Q}}_{\rm quark}  \left(1 + \sum_{n=1}^\infty \frac{g_n}{r^{2n}} + \delta g \, r^{N} \right)  \,, \\
h & = & \frac{h_0}{\hat{\mathcal{Q}}_{\rm quark}^{9/2} r^9} \left(1 + \sum_{n=1}^\infty \frac{h_n}{r^{2n}} + \delta h \, r^{N}\right)  \,, \\
\Phi & = & 2 \log (\sqrt{\hat{\mathcal{Q}}_{\rm quark}} r)+\log \left( \frac{1}{10} \sqrt{\frac{7}{2}} \right) + \sum_{n=1}^\infty \frac{p_n}{r^{2n}}  + \delta \phi \, r^{N}\,. \label{eq:IRsoln}
\eea
The coefficients $\{f_n, g_n, h_n, p_n\}$ are uniquely determined by the equations of motion, while the perturbation $\{\delta f, \partial g, \partial h, \partial \phi \}$ has an overall free magnitude. This irrelevant perturbation has
\beq
N = -\frac{3}{2} \left(\sqrt{79}-3\right) < 0 \,.
\eeq
Different magnitudes for the perturbation will map onto different values of the dimensionless ratio of relevant couplings $\phi_0/\hat{\mu}$  in the UV theory, where $\hat{\mu}$ is the boundary chemical potential rescaled by appropriate factors of $\kappa,e$ and $L$.

We solve the equations of motion, (\ref{eq:EMDF}), by numerically integrating up the IR solution, (\ref{eq:IRsoln}), to the UV boundary at small values of $r$. The only free parameter is the strength of the irrelevant perturbation in the IR. We can read off the relevant coupling $\phi_0$ near the boundary by fitting to the boundary expansion of $\Phi$ (\ref{phiUV}). Similarly, the boundary chemical potential, $\hat{\mu}$, and the total charge, $\hat{\mathcal{Q}}$, can be found from fitting to the boundary expansion of $h(r)=c (\hat{\mu}-\hat{\mathcal{Q}} r)+\dots$. 
A key feature of this IR-UV matching procedure is that the several parameters characterizing the UV limit to do not modify the IR limit of
$g$, and hence do not influence the $\mathcal{Q}$ dependence of the entanglement entropy, as we have also emphasized in Section~\ref{sec:intro}.
In performing the integration we must set $\{\hat{\sigma}, \hat{\rho}, \hat{p}\}$ to zero whenever $\hat{\mu}_{\rm loc}(r)<\hat{m}$ and keep them in the equations otherwise.

\begin{figure}[tbp]
  \centering
  \includegraphics[width=4in]{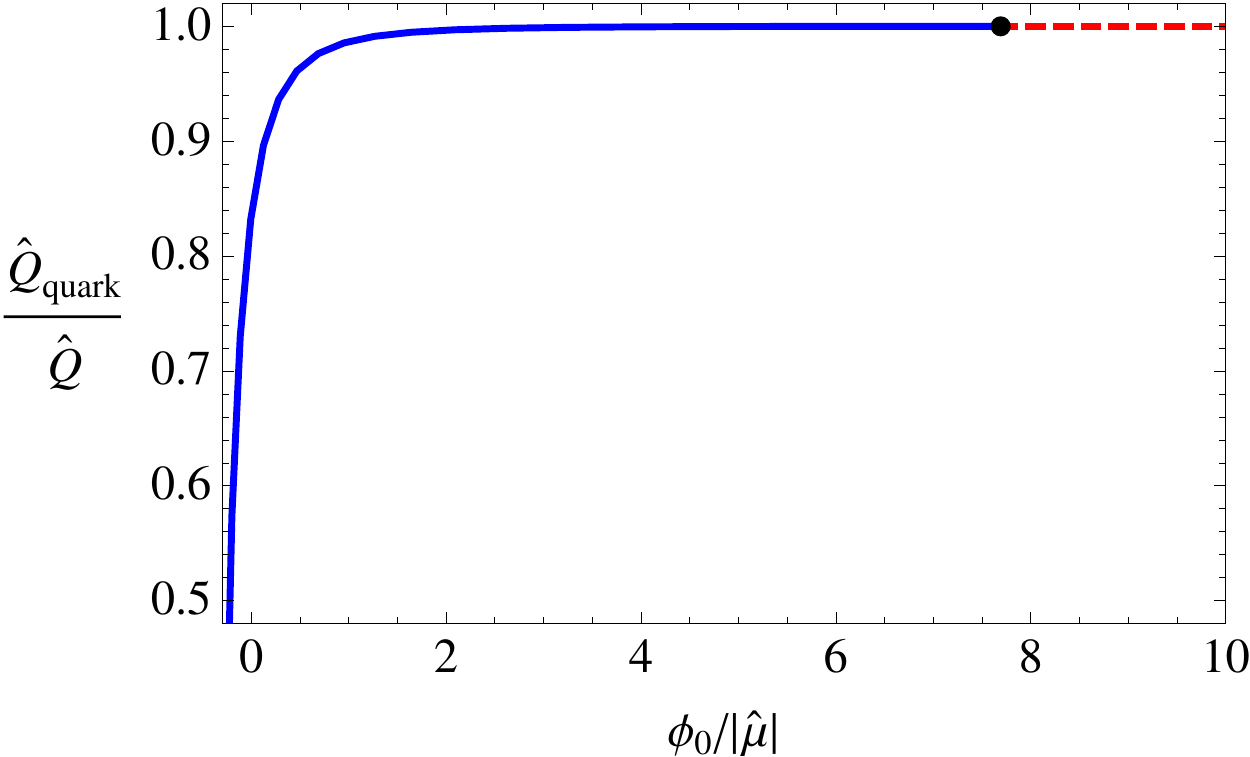}
  \caption{Ratio of fractionalized charge to total charge $\hat{\mathcal{Q}}_{\rm quark}/\hat{\mathcal{Q}}$ as a function of the relevant coupling $\phi_0$ for $\{\hat{m}, \hat{\beta}\} = \{3/16, 10 \}$. The black dot denotes the location of the transition between the FL* and NFL phases. In the FL* phase indicated by the blue drawn line $\hat{\mathcal{Q}}_{\rm quark}/\hat{\mathcal{Q}}<1$, whereas in the NFL phase indicated by the dashed red line we have $\hat{\mathcal{Q}}_{\rm quark}/\hat{\mathcal{Q}}=1$.
  }
  \label{fig:Qfrac}
\end{figure}

In figure \ref{fig:Qfrac}, we plot the ratio of the charge carried by the quarks and the total charge as a function of the relevant coupling, $\phi_0$. As in \cite{lizasean} there is a third order transition between a partially fractionalized phase and a fully fractionalized phase. In the fully fractionalized phase we have $\hat{\mu}_{\rm loc}(r)<\hat{m}$ throughout the entire bulk. It follows that the total charge $\mathcal{Q} = \mathcal{Q}_{\rm quark}$. In the partially fractionalized phase the fluid is present in a region between $r_1$ and $r_2$ in the bulk. In this phase both quarks and mesinos contribute to the total charge: $\mathcal{Q} = \mathcal{Q}_{\rm quark}+ \mathcal{Q}_{\rm mesino}$.

We have argued throughout this paper that for the present Einstein-Maxwell-dilaton-fluid theory the holographic entanglement entropy indicates the presence of hidden Fermi surfaces of gauge-charged quarks in both phases. Furthermore, the presence of Fermi surfaces of gauge-neutral mesinos associated with the fluid was established in \cite{sean2}. We can thus identify this transition with a phase transition between FL* and NFL. It is clear from (\ref{eq:IRsoln}) that the entanglement entropy in the FL* phase is also given by (\ref{see}), but with $\mathcal{Q} \to \mathcal{Q}_{\rm quark}$. Via the Luttinger relation we find that this is consistent with the interpretation of the holographic entanglement entropy being a measure of the hidden quark Fermi surfaces.

\begin{figure}[tbp]
  \centering
  \includegraphics[width=4in]{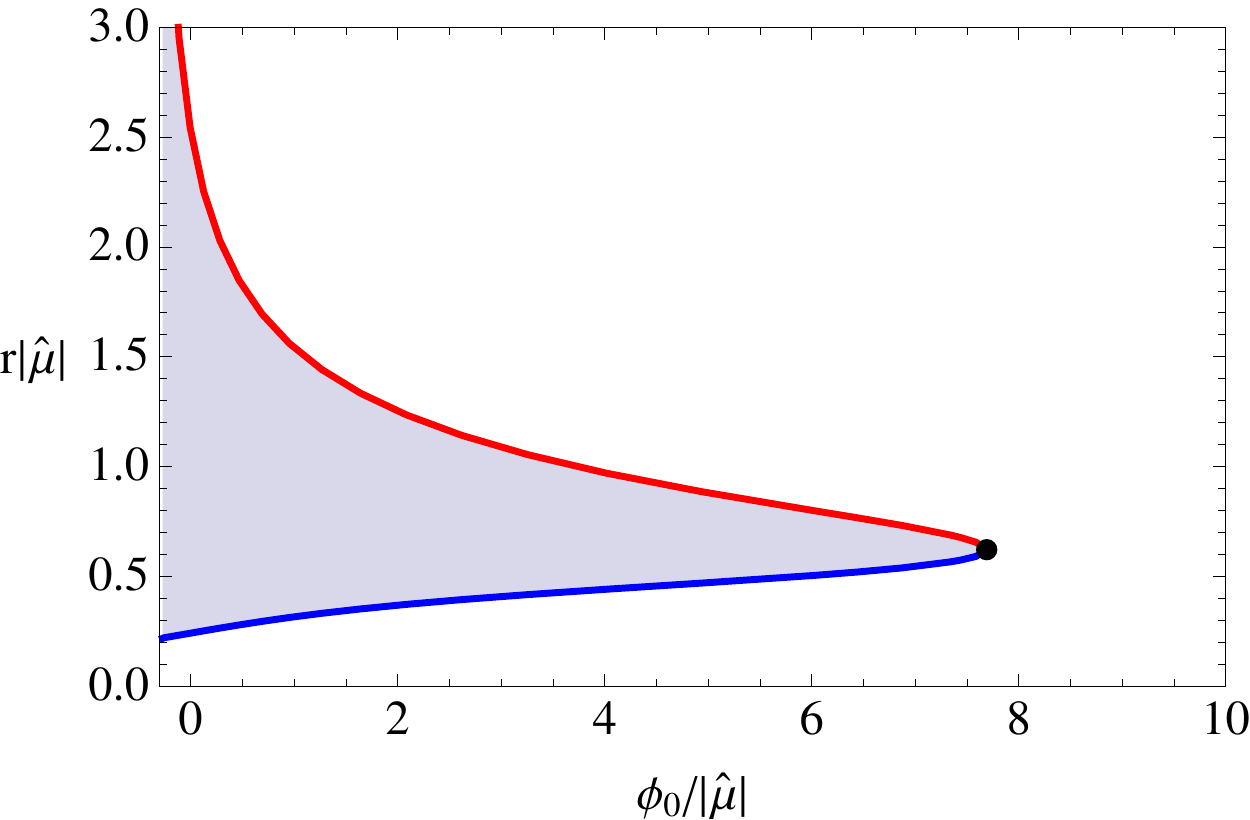}
  \caption{The plot shows the minimal and maximal radii for which the fluid is present as a function of the relevant coupling $\phi_0$. The factors of $\hat{\mu}$ ensure that all quantities are given in dimensionless units. Remember that the IR corresponds to $r \to \infty$. The black dot denotes the location of the transition between the FL* and NFL phases. The fluid is present at all radii for which $\hat{\mu}_{\rm loc}(r)>\hat{m}$, this is indicated by the shaded region; this corresponds to the shaded region
  in Fig.~\ref{fig:ffl}. It follows from scaling arguments that the fluid is present at $r \sim \sqrt{\mathcal{Q}_{\rm quark}}$.
  }
  \label{fig:fluid}
\end{figure}

In figure \ref{fig:fluid} we show the region in the bulk where the fluid is present; see also Fig.~\ref{fig:ffl}. It is clear that this region shrinks to zero at the FL* to NFL transition. It follows from the scaling transformations (\ref{Qscale}) that $\hat{\mu}_{\rm loc}(r)=
F(\mathcal{Q}_{\rm quark}r^2)$. From this we conclude that the fluid is present at $r \sim \sqrt{\mathcal{Q}_{\rm quark}}$.
The discussion below (\ref{chiscale}), applied to the present context, implies that the UV matching conditions do not modify
this conclusion.

For completeness we mention that as in \cite{lizasean} the present theory has two more IR solutions. These solutions describe the phase diagram beyond the FL* phase. There is a Lifshitz solution where the dilaton is constant. This solution can be reached from the UV by fine tuning the coupling to the relevant operator, $\phi_0$, and bounds the partially fractionalized phase. Finally, there is a solution with $\Phi \to -\infty$ in the IR. In \cite{lizasean} this was a domain wall-like solution plus a fluid corresponding to a phase where the flux vanishes in the IR and all the charge is accounted for by mesino Fermi surfaces. For the present theory, however, the solution is quite different. In particular, the fluid density diverges in the IR and so does the flux. We leave the study of this phase for future work. We just mention that this phase is not present if one chooses $Z$ symmetric under $\Phi \to - \Phi$, $Z(\Phi)=Z_0 \cosh(\alpha \Phi)$ for instance.

\section{Discussion}
\label{sec:conc}

We conclude by discussing some issues raised by our results for future work.

The holographic entanglement entropy formula of Ryu-Takayanagi \cite{rt} appears to detect
only the hidden Fermi surfaces. The visible Fermi surfaces of mesinos in our work, and in previous work \cite{sslee0,hong0,zaanen1,hong1,sean2,hong4,ssfl,lizasean}, do properly reduce the net charge associated with the hidden
Fermi surfaces, but do not contribute themselves to the entanglement associated with the  Ryu-Takayanagi result. Presumably,
their contributions will appear in fluctuation contributions to the entanglement entropy: it would be useful to sort this out in future work.

We have analyzed zero temperature solutions of the EMD theory that possess logarithmic violations of the boundary law for entanglement entropy.  We now make a few additional comments on the singularity structure of these solutions along the lines of Ref.~\cite{horowitz} (see also Refs.~\cite{kiritsis,kiritsis2}).  We focus on the case of $d=2$ for concreteness, but our remarks apply more generally.  As a preliminary, we note that the components of the curvature tensor are all finite in the coordinate system of (\ref{metric}) for $f \sim r^{-4(z-1)}$ and $g \sim r^2$ as in (\ref{zmetric}).  However, the spacetime is not geodesically complete.  Consider a radial geodesic with tangent vector $n = (\dot{t},\dot{r},0,0)$ where $\dot{r}$ indicates a derivative with respect to the proper time $\tau$.  By imposing the normalization condition $n \cdot n = -1$ and using the existence of a conserved ``energy" $ n \cdot \partial_t $ we can find a simple differential equation for $r(\tau)$.  Omitting the details, we find an equation of the form
\beq
\dot{r}^2 = \frac{r^2_0}{\tau_0^2} \left(\frac{r^{4z-2}}{ r^{4z-2}_0} -1 \right)
\eeq
with $r_0$ and $\tau_0$ some constants.  For large $r$ this equation may be integrated from $\tau_i$ to yield
\beq
\frac{-1}{2z-2} \left(\frac{r_0^{2z-2}}{r^{2z-2}(\tau)} - \frac{r_0^{2z-2}}{r^{2z-2}(\tau_i)}\right) =  \frac{\tau - \tau_i}{\tau_0}.
\eeq
This geodesic reaches $r=\infty$ in finite proper time $\tau_\infty$ given by
\beq
\frac{1}{2z-2} \frac{r_0^{2z-2}}{r^{2z-2}(\tau_i)} = \frac{\tau_\infty - \tau_i}{\tau_0}
\eeq
and hence the spacetime is not geodesically complete for $z>1$.  The inequality (1.9) with $\theta = d-1$ guarantees that our solutions always have this finite proper time singularity.  We also expect to develop tidal singularities analogous to those in \cite{horowitz} in a parallel transported frame.  These observations have practical importance in that we should apply absorbing boundary conditions at $r=\infty$ when computing correlation functions.

We also have perfectly sensible finite temperature solutions that still display a logarithmic violation of the boundary law for low enough temperatures. One possible interpretation of this situation is to regard the singular zero temperature solution as a kind of intermediate scale description of the physics that, however, may not be extendible all the way to zero temperature.  A small finite temperature provides a regulator and these solutions still describe physics in a wide range of energy scales.  In Ref.~\cite{tadashi}, various other spatial metrics were written (without full solutions) that give a greater than logarithmic violation of the boundary law.  From the perspective of field theory, these solutions are quite anomalous in terms of their entanglement entropy, so like the extremal Reissner-N\"ordstrom
black hole, we believe that such a description must not be valid to arbitrarily low energies.  As argued in Ref.~\cite{horowitz},
such a breakdown is visible on the gravity side in that the deep IR geometry tends to excite even very heavy bulk degrees of freedom.

There is an important distinction between these more anomalous states and the hidden Fermi surfaces we have considered: the hidden Fermi surface system can be stable to zero temperature.  In fact, this stability is a non-trivial question on the field theory side especially given the propensity of the Fermi liquid to form paired superfluid states.  For a Fermi surface coupled to a U(1) gauge field, there is a range of parameters where the system is stable to pairing \cite{max}.  However, the situation for a Fermi surface coupled to a non-Abelian gauge field, as may be more appropriate in the EMD context, is less favorable.  The colored ``quarks" may be paired into a wider variety of color states owing to the non-Abelian nature of the group, and invariably at least one of these channels is attractive.
Whether this conclusion can be modified by the inclusion of other strongly coupled IR degrees of freedom, such as those that appear in the EMD theory, is not known.  Thus although there remain issues associated with the singularity structure of our solutions, the physics they appear to describe is compatible with a true zero temperature compressible state.  In the future, it would be interesting to develop additional bulk criteria that can explicitly rule out the anomalous solutions.

We mention one final interesting point.  Free fermions and Fermi liquids are also known to have anomalous fluctuations in conserved charges \cite{klich,swingleMI}. Such fluctuations are interesting in part because of their greater experimental accessibility.  Let $Q_A = \int_A d^d x \, \mathcal{Q}$ be the charge operator restricted to region $A$ of linear size $R$.  Even if the total charge $Q$ commutes with the Hamiltonian, the restricted operator $Q_A$ need not.  It can be shown that CFTs with a conserved charge have $\langle Q_A \rangle =0$ and fluctuations $\Delta Q_A^2 = \langle Q_A^2 \rangle \sim R^{d-1}$ ($d>1$).  The compressible Fermi systems mentioned above have instead $\langle Q_A \rangle \neq 0$ and $\Delta Q_A^2 = \langle (Q_A - \langle Q_A \rangle)^2 \rangle \sim R^{d-1} \log{(R)}$.  We suspect similar charge fluctuations will be present in general compressible states, so it would be quite interesting to perform a holographic calculation of these fluctuations.

\acknowledgements

We thank T.~Grover, S.~Kachru, E.~Kiritsis, H.~Liu, M.~Metlitski, 
T.~Takayanagi, S.~Trivedi, and especially S.~Hartnoll for valuable discussions. This research was supported by the National Science Foundation under grants DMR-1103860 and PHY05-51164, by a MURI grant from AFOSR. LH acknowledges funding from the Netherlands Organisation for Scientific Research (NWO).
BS is supported by a Simons Fellowship through Harvard University.


\begin{thebibliography}{}

\bibitem{nernst}
  S.~A.~Hartnoll, P.~K.~Kovtun, M.~M\"uller, and S.~Sachdev,
  ``Theory of the Nernst effect near quantum phase transitions in condensed matter, and in dyonic black holes,''
  Phys.\ Rev.\  {\bf B76}, 144502 (2007)
  [arXiv:0706.3215 [cond-mat.str-el]].

\bibitem{sslee0}
  S.-S.~Lee,
  ``A Non-Fermi Liquid from a Charged Black Hole: A Critical Fermi Ball,''
  Phys.\ Rev.\  D {\bf 79}, 086006 (2009)
  [arXiv:0809.3402 [hep-th]].

\bibitem{denef0}
F.~Denef and S.~A.~Hartnoll,
  ``Landscape of superconducting membranes,''
  Phys.\ Rev.\ D {\bf 79}, 126008 (2009)
  [arXiv:0901.1160 [hep-th]].

\bibitem{hong0}
  H.~Liu, J.~McGreevy, and D.~Vegh,
  ``Non-Fermi liquids from holography,''
  arXiv:0903.2477 [hep-th].

  \bibitem{zaanen1}
  M.~\v{C}ubrovi\'{c}, J.~Zaanen, and K.~Schalm,
  ``String Theory, Quantum Phase Transitions and the Emergent Fermi-Liquid,''
  Science {\bf 325}, 439 (2009).
  [arXiv:0904.1993 [hep-th]].

\bibitem{hong1} T.~Faulkner, H.~Liu, J.~McGreevy, and D.~Vegh,
  ``Emergent quantum criticality, Fermi surfaces, and AdS(2),''
  arXiv:0907.2694 [hep-th].

\bibitem{denef}
  F.~Denef, S.~A.~Hartnoll, and S.~Sachdev,
  ``Quantum oscillations and black hole ringing,''
  Phys.\ Rev.\ D {\bf 80}, 126016 (2009)
  [arXiv:0908.1788 [hep-th]].

\bibitem{faulkner}  T.~Faulkner and J.~Polchinski,
  ``Semi-Holographic Fermi Liquids,''
  JHEP {\bf 1106}, 012 (2011)
  [arXiv:1001.5049 [hep-th]].

\bibitem{pufuigor}
  C.~P.~Herzog, I.~R.~Klebanov, S.~S.~Pufu and T.~Tesileanu,
  ``Emergent Quantum Near-Criticality from Baryonic Black Branes,''
  JHEP\ {\bf 1003}, 093  (2010)
  [arXiv:0911.0400 [hep-th]].

\bibitem{polchinski}  S.~A.~Hartnoll, J.~Polchinski, E.~Silverstein, and D.~Tong,
  ``Towards strange metallic holography,''
  JHEP {\bf 1004}, 120 (2010)
  [arXiv:0912.1061 [hep-th]].

\bibitem{gubserrocha}    S.~S.~Gubser and F.~D.~Rocha,
  ``Peculiar properties of a charged dilatonic black hole in AdS$_5$,''
  Phys.\ Rev.\  D {\bf 81}, 046001 (2010)
  [arXiv:0911.2898 [hep-th]].

\bibitem{hong2}
  T.~Faulkner, N.~Iqbal, H.~Liu, J.~McGreevy, and D.~Vegh,
  ``Strange metal transport realized by gauge/gravity duality,''
  Science {\bf 329}, 1043 (2010).

\bibitem{kiritsis}   C.~Charmousis, B.~Gouteraux, B.~S.~Kim, E.~Kiritsis, and R.~Meyer,
  ``Effective Holographic Theories for low-temperature condensed matter systems,''
  arXiv:1005.4690 [hep-th].
  
\bibitem{kiritsis2}
  B.~Gouteraux and E.~Kiritsis,
  ``Generalized Holographic Quantum Criticality at Finite Density,''
  arXiv:1107.2116 [hep-th].
  
\bibitem{sean1}   S.~A.~Hartnoll and A.~Tavanfar,
  ``Electron stars for holographic metallic criticality,''
  Phys.\ Rev.\ D {\bf 83}, 046003 (2011)
  [arXiv:1008.2828 [hep-th]].

\bibitem{sean2}    S.~A.~Hartnoll, D.~M.~Hofman and A.~Tavanfar,
  ``Holographically smeared Fermi surface: Quantum oscillations and Luttinger count in electron stars,''
  Europhys.\ Lett.\  {\bf 95}, 31002 (2011)
  [arXiv:1011.2502 [hep-th]].

\bibitem{sean3}   S.~A.~Hartnoll, D.~M.~Hofman and D.~Vegh,
  ``Stellar spectroscopy: Fermions and holographic Lifshitz criticality,''
  JHEP {\bf 1108}, 096 (2011)
  [arXiv:1105.3197 [hep-th]].

\bibitem{seanr}   S.~A.~Hartnoll,
  ``Horizons, holography and condensed matter,''
  arXiv:1106.4324 [hep-th].

\bibitem{larus1}   E.~J.~Brynjolfsson, U.~H.~Danielsson, L.~Thorlacius, and T.~Zingg,
  ``Black Hole Thermodynamics and Heavy Fermion Metals,''
  JHEP {\bf 1008}, 027 (2010)
  [arXiv:1003.5361 [hep-th]].

\bibitem{larus2}   V.~G.~M.~Puletti, S.~Nowling, L.~Thorlacius, and T.~Zingg,
  ``Holographic metals at finite temperature,''
  JHEP {\bf 1101}, 117 (2011)
  [arXiv:1011.6261 [hep-th]].

\bibitem{eric} X.~Arsiwalla, J.~de Boer, K.~Papadodimas, and E.~Verlinde,
  ``Degenerate Stars and Gravitational Collapse in AdS/CFT,''
  arXiv:1010.5784 [hep-th].

\bibitem{kachru2}
    K.~Goldstein, S.~Kachru, S.~Prakash, and S.~P.~Trivedi,
  ``Holography of Charged Dilaton Black Holes,''
  JHEP {\bf 1008}, 078 (2010)
  [arXiv:0911.3586 [hep-th]].

\bibitem{kachru3}
    K.~Goldstein, N.~Iizuka, S.~Kachru, S.~Prakash, S.~P.~Trivedi, and A.~Westphal,
  ``Holography of Dyonic Dilaton Black Branes,''
  JHEP {\bf 1010}, 027 (2010)
  [arXiv:1007.2490 [hep-th]].

\bibitem{kachru4}    K.~Jensen, S.~Kachru, A.~Karch, J.~Polchinski and E.~Silverstein,
  ``Towards a holographic marginal Fermi liquid,''
  Phys.\ Rev.\ D {\bf 84}, 126002 (2011)
  [arXiv:1105.1772 [hep-th]].


\bibitem{trivedi}   N.~Iizuka, N.~Kundu, P.~Narayan, and S.~P.~Trivedi,
  ``Holographic Fermi and Non-Fermi Liquids with Transitions in Dilaton Gravity,''
  arXiv:1105.1162 [hep-th].

\bibitem{zaanen2}  M.~Cubrovic, J.~Zaanen and K.~Schalm,
  ``Constructing the AdS dual of a Fermi liquid: AdS Black holes with Dirac hair,''
  JHEP {\bf 1110}, 017 (2011)
  [arXiv:1012.5681 [hep-th]].


\bibitem{ssffl}   S.~Sachdev,
  ``Holographic metals and the fractionalized Fermi liquid,''
  Phys.\ Rev.\ Lett.\  {\bf 105}, 151602 (2010)
  [arXiv:1006.3794 [hep-th]].

\bibitem{mcphys} J.~McGreevy, ``In pursuit of a nameless metal,'' Physics {\bf 3}, 83 (2010).

\bibitem{liza} L.~Huijse and S.~Sachdev, ``Fermi surfaces and gauge-gravity duality,''
Phys. Rev. D {\bf 84}, 026001 (2011) [arXiv:1104.5022 [hep-th]].

\bibitem{leiden}    M.~Cubrovic, Y.~Liu, K.~Schalm, Y.~-W.~Sun and J.~Zaanen,
  ``Spectral probes of the holographic Fermi groundstate: dialing between the electron star and AdS Dirac hair,''
  Phys.\ Rev.\ D {\bf 84}, 086002 (2011)
  [arXiv:1106.1798 [hep-th]].

\bibitem{hong4}   N.~Iqbal, H.~Liu, and M.~Mezei,
  ``Semi-local quantum liquids,''
  arXiv:1105.4621 [hep-th].

\bibitem{hong5} 
  N.~Iqbal, H.~Liu and M.~Mezei,
  ``Lectures on holographic non-Fermi liquids and quantum phase transitions,''
  arXiv:1110.3814 [hep-th].

\bibitem{pp} M.~Edalati, K.~W.~Lo, and P.~W.~Phillips,
  ``Neutral Order Parameters in Metallic Criticality in $d=2+1$ from a Hairy Electron Star,''
  arXiv:1106.3139 [hep-th].

\bibitem{waldram}   J.~P.~Gauntlett, J.~Sonner and D.~Waldram,
  ``Universal fermionic spectral functions from string theory,''
  Phys.\ Rev.\ Lett.\  {\bf 107}, 241601 (2011)
  [arXiv:1106.4694 [hep-th]].

\bibitem{yarom}   R.~Belliard, S.~S.~Gubser and A.~Yarom,
  ``Absence of a Fermi surface in classical minimal four-dimensional gauged supergravity,''
  JHEP {\bf 1110}, 055 (2011)
  [arXiv:1106.6030 [hep-th]].


\bibitem{ssfl} S.~Sachdev,   ``A model of a Fermi liquid using gauge-gravity duality,''
  Phys.\ Rev.\  D {\bf 84}, 066009 (2011)
  [arXiv:1107.5321 [hep-th]].

\bibitem{ssarcmp}  S.~Sachdev,
  ``What can gauge-gravity duality teach us about condensed matter physics?,''
  to appear in Annual Reviews of Condensed Matter Physics,
  arXiv:1108.1197 [cond-mat.str-el].

\bibitem{larus3}   V.~G.~M.~Puletti, S.~Nowling, L.~Thorlacius, and T.~Zingg,
  ``Friedel Oscillations in Holographic Metals,''
  arXiv:1110.4601 [hep-th].

\bibitem{lizasean}    S.~A.~Hartnoll and L.~Huijse,
  ``Fractionalization of holographic Fermi surfaces,''
  arXiv:1111.2606 [hep-th].

\bibitem{tadashi}   N.~Ogawa, T.~Takayanagi, and T.~Ugajin,
  ``Holographic Fermi Surfaces and Entanglement Entropy,''
  arXiv:1111.1023 [hep-th].

\bibitem{ffl1} T.~Senthil, S.~Sachdev, and M.~Vojta, ``Fractionalized Fermi liquids,''
Phys. Rev. Lett. {\bf 90}, 216403 (2003)
[arXiv:cond-mat/0209144].

\bibitem{ffl2} T.~Senthil, M.~Vojta, and S.~Sachdev, ``Weak magnetism and non-Fermi liquids near heavy-fermion
critical points,'' Phys. Rev. B {\bf 69}, 035111 (2004)
[arXiv:cond-mat/0305193].

\bibitem{swingle} B. Swingle, ``Entanglement Entropy and the Fermi Surface,''
Phys. Rev. Lett. {\bf 105}, 050502 (2010) [arXiv:0908.1724 [cond-mat.str-el]].

\bibitem{tarun} Y.~Zhang, T.~Grover, and A.~Vishwanath,
``Entanglement entropy of critical spin liquids,'' Phys. Rev. Lett. {\bf 107} 067202
(2011) [arXiv:1102.0350 [cond-mat.str-el]].

\bibitem{seidel} W.~Ding, A.~Seidel, and K.~Yang, ``Entanglement Entropy of Fermi Liquids via Multi-dimensional Bosonization,''
arXiv:1110.3004 [cond-mat.stat-mech].

\bibitem{motfish} O.~I~ Motrunich and M.~P.~A.~Fisher, ``D-wave correlated Critical Bose Liquids in two dimensions,''
Phys. Rev. B {\bf 75}, 235116 (2007) [arXiv:cond-mat/0703261].

\bibitem{mulligan} S.~Kachru, X.~Liu, and M.~Mulligan, ``Gravity Duals of Lifshitz-like Fixed Points,''
Phys. Rev. D {\bf 78}, 106005 (2008). [arXiv:0808.1725 [hep-th]].

\bibitem{horowitz} G.~Horowitz and B.~Way, ``Lifshitz singularities,'' arXiv:1111.1243 [hep-th].

\bibitem{dsf} D.~S.~Fisher, ``Scaling and critical slowing down in random-field Ising systems,'' Phys. Rev. Lett. {\bf 56}, 416 (1986).

\bibitem{fwgf} M.~P.~A.~Fisher, P.~B.~Weichman, G.~Grinstein, and D.~S.~Fisher,
``Boson localization and the superfluid-insulator transition,''
Phys.\ Rev.\ B {\bf 40}, 546 (1989).

\bibitem{book} S.~Sachdev, {\em Quantum Phase Transitions},
2nd Edition, Cambridge University Press (2011).

\bibitem{leen} Sung-Sik Lee, ``Low energy effective theory of Fermi surface coupled with U(1) gauge field in 2+1 dimensions,''
Phys. Rev. B {\bf 80}, 165102 (2009)  [arXiv:0905.4532 [cond-mat.str-el]].

\bibitem{metnem}  M.~A.~Metlitski, and S.~Sachdev,
  ``Quantum phase transitions of metals in two spatial dimensions: I. Ising-nematic order,''
  Phys.\ Rev.\  B {\bf 82}, 075127 (2010)
  [arXiv:1001.1153 [cond-mat.str-el]].

\bibitem{mross} D.~F.~Mross, J.~McGreevy, H.~Liu, and T.~Senthil,
  ``A controlled expansion for certain non-Fermi liquid metals,''
  Phys. Rev. B {\bf 82}, 045121 (2010)
  [arXiv:1003.0894 [cond-mat.str-el]].

\bibitem{rt} S.~Ryu and T.~Takayanagi,
  ``Holographic derivation of entanglement entropy from AdS/CFT,''
  Phys.\ Rev.\ Lett.\  {\bf 96}, 181602 (2006)
  [arXiv:hep-th/0603001].

\bibitem{cardy} P. Calabrese and J. L. Cardy, ``Entanglement entropy and quantum field theory,''
J. Stat. Mech. {\bf 0406}, P002 (2004) [arXiv:hep-th/0405152].

\bibitem{mfs} M.~A.~Metlitski, C.~A.~Fuertes, and S.~Sachdev, ``Entanglement Entropy in the O($N$) model,''
Phys.\ Rev.\ B {\bf 80}, 115122 (2009) [arXiv:0904.4477 [cond-mat.stat-mech]].

\bibitem{klich} D.~Gioev and I.~Klich, ``Entanglement entropy of fermions in any dimension and the Widom conjecture,''
Phys. Rev. Lett. {\bf 96}, 100503 (2006) [arXiv:quant-ph/0504151].

\bibitem{wolf} M.~M.~Wolf, ``Violation of the entropic area law for Fermions,''
Phys. Rev. Lett. {\bf 96}, 010404 (2006) [arXiv:quant-ph/0503219].

\bibitem{barthel}  T.~Barthel, M.-C.~Chung, and U.~Schollwock,
``Entanglement scaling in critical two-dimensional fermionic and bosonic systems,''
Phys. Rev. A {\bf 74}, 022329 (2006) [arXiv:cond-mat/0602077].

\bibitem{haas1} W. Li, L. Ding, R. Yu, T. Roscilde, and S. Haas,
   ``Scaling Behavior of Entanglement in Two- and Three-Dimensional Free Fermions,''
   Phys. Rev. B {\bf 74}, 073103 (2006) [arXiv:quant-ph/0602094].

\bibitem{vicari} P.~Calabrese, M.~Mintchev, and E.~Vicari,
``Entanglement entropies in free fermion gases for arbitrary dimension,''
Europhys. Lett. {\bf 97}, 20009 (2012) 
[arXiv:1110.6276 [cond-mat.quant.gas]].

\bibitem{susy}  L.~Huijse and K.~Schoutens, ``Superfrustration of charge degrees of freedom,''
Eur. Phys. J. B {\bf 64}, 543 (2008) arXiv:0709.4120 [cond-mat.str-el]]
and references therein.

\bibitem{metzner}  S.~Thier and W.~Metzner, ``Singular order parameter interaction at nematic quantum critical point in two dimensional electron systems,''
arXiv:1108.1929 [cond-mat.str-el].

\bibitem{sonqcd}   D.~T.~Son,
  ``Superconductivity by long range color magnetic interaction in high density quark matter,''
  Phys.\ Rev.\ D\ {\bf 59}, 094019  (1999)
  [hep-ph/9812287].

\bibitem{garyreview}   G.~T.~Horowitz,
  ``The dark side of string theory: Black holes and black strings.,''
  Trieste 1992, Proceedings, String theory and quantum gravity, 55-99,
  arXiv:hep-th/9210119.

\bibitem{myers} H.~Casini, and M.~Huerta, and R.~Myers, ``Towards a derivation of holographic entanglement entropy,'' arXiv:1102.0440 [hep-th].

\bibitem{swingle2} B.~Swingle and T.~Senthil, ``Universal crossovers between entanglement entropy and thermal entropy,'' 
arXiv:1112.1069 [cond-mat.str-el].

\bibitem{swingleMI} B.~Swingle, ``Renyi entropy, mutual information, and fluctuation properties of Fermi liquids,'' arXiv:1007.4825 [cond-mat.str-el].

\bibitem{max} M.~Metlitski, D.~Mross, S.~Sachdev, and T.~Senthil, Bull. Am. Phys. Soc. {\bf 57}, No. 1, W16.00001
(2012), http://meetings.aps.org/Meeting/MAR12/Event/166339.

\end{thebibliography}
\end{document}